\documentclass[11pt]{article}
\usepackage{cite}
\usepackage{amsmath,amsfonts,amssymb}
\usepackage[small,bf,hang]{caption}
\usepackage{slashed}
\usepackage{color}

\def\hybrid{
        \topmargin -20pt
        \oddsidemargin 0pt
        \headheight 0pt \headsep 0pt
        \textwidth 6.25in 
        \textheight 9.5in 
        \marginparwidth .875in
        \parskip 5pt plus 1pt \jot = 1.5ex}

\hybrid

 \csname
@addtoreset\endcsname{equation}{section}

\newcommand{\qeq}{{\hbox{=\kern-2.3mm ? \kern.5mm }}}
\renewcommand{\qeq}{=}

\newcommand{\HH}{{\cal H}}

\newcommand{\CC}{{\cal C}}

\newcommand{\wt}{\widetilde}
\newcommand{\wh}{\widehat}

\newcommand{\ba}{\bar a}

\newcommand{\be}{\begin{equation}}
\newcommand{\ee}{\end{equation}}
\newcommand{\ben}{\begin{eqnarray}\displaystyle}
\newcommand{\een}{\end{eqnarray}}

\newcommand{\refb}[1]{(\ref{#1})}
\newcommand{\p}{\partial}
\newcommand{\sectiono}[1]{\section{#1}\setcounter{equation}{0}}

\def\moth{\mathsurround=0pt}
\newdimen\zo \zo=0pt

\def\tick{\leaders\hrule height 0.5ex depth 0pt \hskip 0.5pt}
\def\upboxfill{$\moth \setbox\zo\hbox{\tick}%
  \hskip 3pt\hbox to 0pt{$\tick$\hss}\hrulefill \hbox to 7.5pt{$\tick$\hss}$}

\def\dtick{\leaders\hrule height .34pt depth 0.5ex \hskip 0.5pt}
\def\downboxfill{$\moth \setbox\zo\hbox{\dtick}%
  \hskip 2pt\hbox to 0pt{$\dtick$\hss}\hrulefill \hbox to 2pt{$\dtick$\hss}$}

\def\bec{\begin{center}}
\def\ec{\end{center}}

 \def\det{{\rm det\,}}
\def\be{\begin{equation}}
\def\ee{\end{equation}}
\def\bea{\begin{eqnarray}}
\def\eea{\end{eqnarray}}
\def\ba{\begin{array}}
\def\ea{\end{array}}

\begin{document}

\begin{titlepage}
\rightline{November 2014} 
\rightline{\tt MIT-CTP-4603} 
\rightline{\tt HRI/ST/1415} 
\begin{center}
\vskip 2.5cm

{\Large \bf { Heterotic Effective Action and Duality Symmetries Revisited }}\\

 \vskip 2.0cm
{\large {Olaf Hohm${}^1$, Ashoke Sen${}^2$ and Barton Zwiebach${}^1$}}
\vskip 0.5cm
{\it {${}^1$Center for Theoretical Physics}}\\
{\it {Massachusetts Institute of Technology}}\\
{\it {Cambridge, MA 02139, USA}}\\[2.5ex]
{\it {${}^2$Harishchandra Research Institute}}\\
{\it {Chhatnag Road, Jhusi}}\\
{\it {Allahabad 211019, India}}\\
ohohm@mit.edu, sen@hri.res.in, zwiebach@mit.edu\\   

\vskip 2.5cm
{\bf Abstract}

\end{center}

\vskip 0.5cm

\noindent
\begin{narrower}

\baselineskip15pt

The dimensional  
reduction of heterotic supergravity 
with gauge fields truncated to the Cartan subalgebra exhibits a continuous 
$O(d,d+16;\mathbb{R})$ global symmetry, related to the $O(d,d+16;\mathbb{Z})$ T-duality of heterotic strings on a $d$-torus.
 The $O(d,d+16;\mathbb{R})$ symmetry is not present, however, if the 
supergravity reduction is done
including the full set of $E_8\times E_8$ or $SO(32)$ gauge fields. 
We analyze which duality symmetries are realized to all orders in 
$\alpha'$ in the proper effective field theories for the 
massless string states.
We find a universal $O(d,d;\mathbb{R})$ symmetry, also predicted by  double field theory. We confirm this by giving a novel formulation of the dimensionally reduced supergravity in terms of 
$O(d,d)$ multiplets, and we discuss cases of symmetry enhancement.

\end{narrower}

\end{titlepage}

\setcounter{tocdepth}{1}
\tableofcontents

\section{Introduction}

The T-duality properties of heterotic string theory were studied 
in the seminal works of Narain~\cite{Narain:1985jj} and
Narain, Sarmadi, and Witten \cite{Narain:1986am}.  The T-duality group $O(d,d+16;\mathbb{Z})$ arises from a compactification on a $d$-dimensional
torus that includes Wilson lines in the Cartan subgroup of the gauge group. 
Duality symmetries have a counterpart in continuous global 
symmetries of the
low-energy action for the massless fields~\cite{Veneziano:1991ek,Sen:1991zi,9109038}.  A particularly clear discussion of this relationship was given by Maharana and Schwarz in~\cite{Maharana:1992my}.  
(For earlier results see \cite{Duff:1985cm,Duff:1986ya}.)
In order to explain the global
$O(d,d+16;\mathbb{R})$ of the low-energy limit of compactified heterotic strings they considered heterotic  supergravity with the gauge group
{\em truncated} to the maximal Cartan subgroup.  They performed  dimensional reduction and displayed the expected global symmetry of the reduced theory.  
With the development 
of double field theory formulations 
\cite{Siegel:1993th,Siegel:1993bj,Hull:2009mi,Hohm:2010jy,Hohm:2010pp} 
 of the low-energy limits of string theories, the manifest display of global duality symmetries and the effect of $\alpha'$ corrections 
 is now the subject of renewed interest 
\cite{Hohm:2011ex,Hohm:2011nu,Baraglia,Garcia-Fernandez,Anderson:2014xha,delaOssa:2014cia,Hohm:2013jaa,Hohm:2014eba,Hohm:2014xsa,Bedoya:2014pma,Coimbra:2014qaa,delaOssa:2014msa,Blumenhagen:2014iua}.  
Motivated by this, we revisit here some aspects
of the continuous T-duality symmetry of the heterotic string effective action.  

Maharana and Schwarz (MS) truncate the higher dimensional 
heterotic supergravity theory to the Cartan subgroup {\em before} performing the reduction. In fact,  
the $O(d,d+16;\mathbb{R})$ symmetry is not present upon 
reduction if one includes any {\em non-abelian}
gauge group. 
This is puzzling because, after all, the gauge group in heterotic string theory is 
$E_8\times E_8$ or $SO(32)$.  The main   
goals of this paper are to clarify,
on general string-theoretic grounds, which 
duality symmetry we should expect
for the effective spacetime theory of the massless fields to any order in $\alpha'$, and to  exhibit this symmetry in a manifest form.

We use 
the symmetries of S-matrix elements of massless states to explain that, to all orders in $\alpha'$, the effective action for the massless fields has a universal global $O(d,d;\mathbb{R})$
symmetry.\footnote{We exclude the massless non-abelian fields which could arise {\it e.g.}
when some of the radii take self-dual values. If we keep these
non-abelian fields then the continuous duality symmetry of the tree effective action of massless fields
could be further reduced.}
The arguments, which are an elaboration of those in 
\cite{Sen:1991zi,9109038}, do not preclude possible enhancements
at points of the moduli space with   
reduced sets of massless fields. 
We also perform the   
direct dimensional reduction of the non-abelian action.  
Since  dimensional reduction 
can be viewed physically as compactification on tori without any field dependence on the toroidal coordinates, we verify that the non-abelian gauge field
reduction is {\em not} constrained by flux quantization.\footnote{Due to the commutator terms in non-abelian field strengths, even constant gauge field configurations give rise to fluxes.}   
The resulting action \refb{ms-nonabelian} is similar to that of MS, with new non-abelian
gauge-covariant couplings and potential terms.  
The $O(d,d;\mathbb{R})$ symmetry of this action is not  
manifest (although it is certainly present).    
Introducing Wilson lines 
 for the Cartan gauge fields corresponds to giving expectation values to the scalars that arise from the internal components of the Cartan gauge fields.\footnote{Given a field with a space-time index, 
we will call the {\em internal} components 
those where the index takes value on the compact coordinates.  
We will call the {\em external} 
components those where the index takes value on the non-compact
coordinates.  For a gauge field, for example, the internal components represent scalars of the lower-dimensional theory and the external components comprise a gauge field of the lower-dimensional theory.
}  In doing so,  all scalars and gauge fields arising from the non-Cartan gauge fields acquire masses via a Higgs mechanism.  They must 
 be dropped  to describe the proper effective action, 
 given that we do not include massive Kaluza-Klein modes nor massive string modes.  Restricted to the massless fields, 
 the action now
 becomes the MS action with an enhanced $O(d,d+16;\mathbb{R})$ symmetry.

The general presence of a global $O(d,d;\mathbb{R})$ 
symmetry to lowest order in $\alpha'$ 
is also implied by the double field theory 
formulation of heterotic supergravity 
\cite{Siegel:1993th,Siegel:1993bj,Hohm:2011ex,Hohm:2011nu}. The doubled formulation
uses $O(d, d+K;\mathbb{R})$ multiplets, with $K$ the {\em dimension}  
of the non-abelian gauge group.  
This $O(d, d+K;\mathbb{R})$ group, however, is not a symmetry. As discussed in \cite{Hohm:2011ex}, the actual global symmetry 
depends on the gauge group but contains at least $O(d,d)$.   
Guided by these results 
we cast the dimensionally reduced supergravity, including {\em  all non-abelian gauge fields},  
into an $O(d, d+K;\mathbb{R})$ `covariant' form (see (\ref{hetACTION})).

Since only $O(d,d)$ is always an actual symmetry, it is 
desirable to formulate the theory in terms of 
$O(d,d)$ multiplets.  
Let us stress that this is a non-trivial problem 
because the symmetries are {\em non-linearly} realized. 
As the main technical result of this paper
we present such a formulation. We write the action 
in terms of an $O(d,d)$ valued 
`generalized metric' ${\cal H}$ and a
Lie algebra valued $O(d,d)$ vector ${\cal C}$.  
Specifically, the scalars arising out of the internal components   
of the metric, 2-form-field, and non-abelian gauge vectors,  
denoted by $G$, $B$, and $a$, respectively,  are  
encoded in the following fields: 
 \be
  O(d,d) \;\;\; \text{field content:} \qquad  \;\; {\cal H}_{MN}\;,\quad {\cal C}_{M}{}^{\alpha}\;, 
 \ee 
where $M,N,\ldots$ are fundamental $O(d,d)$ indices and $\alpha,\beta$ denote the 
adjoint gauge group indices. These fields satisfy the following constraints, written in 
matrix notation:
\be \label{epp4sINTRO}
{\cal H}\, \eta\,  {\cal H} \ = \ \eta\,  , \quad (1+{\cal H}\,\eta) 
\, {\cal C} \ = \ 0\, ,
\ee
where $\eta$ is the $O(d,d)$ invariant metric.  The first constraint
simply states the familiar property of the generalized metric, the second
is an $O(d,d)$ covariant constraint on ${\cal C}$.  
Given the first constraint, the second one is in fact a projector condition that cuts half of the degrees of freedom in ${\cal C}$.
The 
fields can be parametrized in terms of $G$, $B$ and $a$. For ${\cal H}$ we find   
\be
\label{h-physicalINTRO}  
{\cal H}\ = \ \begin{pmatrix}
\bar G^{\,-1} & -\bar G^{\,-1}  B\phantom{\Bigl(}\cr
B\,\bar G^{\,-1}   & \ \bar G\ - \ B\,\bar G^{\,-1}\,B
\end{pmatrix}  \,, 
 \ee
while ${\cal C}$ can be written as    
   \be  
      {\cal C} \  = \  \tfrac{1}{2}
   \begin{pmatrix} - \bar {G}^{\, -1} a^T\kappa  \phantom{\Bigl(}\cr -B\, \bar {G}^{\, -1} a^T\kappa + a^T\kappa 
   \end{pmatrix}\;, 
 \ee  
where $\kappa$ denotes the Cartan-Killing metric of the gauge group.   
For ${\cal H}$ this is the familiar form, except that the 
internal metric $G$  
is redefined with a contribution from the internal $E_8\times E_8$ or $SO(32)$ gauge field components: 
 \be\label{redefg} 
   \bar G \ \equiv \ G+\tfrac{1}{2}a^T\kappa a\; . 
 \ee
The action for dimensionally reduced heterotic supergravity in
these new variables is given in (\ref{hetACTIONFINAL}).
We note that the redefinition (\ref{redefg}) 
 is 
 compatible with the findings of refs.~\cite{Bergshoeff:1995cg} and \cite{Serone:2005ge}, 
which determined the Buscher rules for the heterotic theory with a single circle direction and 
found that such a redefinition 
naturally occurs. Interestingly, a redefinition of the type (\ref{redefg}) 
also featured  in \cite{Hull:1986xn}, for reasons seemingly unrelated to T-duality.

\medskip

So far we have discussed two cases. One is heterotic string compactifications 
with non-zero Wilson lines for the 16 gauge fields in the Cartan subalgebra
of the gauge algebra $G$,  resulting in a moduli space  
$O(d,d+16)/O(d) \times O(d+16)$ and $O(d,d+16;\mathbb{Z})$ dualities. 
The other one  represents the case where there is not a single  Wilson line. 
Here the moduli space is $O(d,d)/O(d) \times O(d)$ and we have
$O(d,d;\mathbb{Z})$ dualities. 
The intermediate situation, however, is also of interest.  
Letting superscripts denote
rank, consider a subgroup 
$G^{(r)} \times G^{(16-r)}$ of the rank 16 gauge group $G$.
We can then imagine a compactification with Wilson lines
for the Cartan gauge fields
$U(1)^{(16-r)}$ of the second factor.
The moduli space of such compactification
 is $O(d,d+16-r)/O(d) \times O(d+16-r)$ and the full string duality
 group is $O(d,d+16-r;\mathbb{Z})$. 
In this case the non-Cartan gauge fields of the second factor,
as well as the gauge fields outside the  
$G^{(r)}\times G^{(16-r)}$ subgroup, acquire masses. 
The massless fields, apart from those from the gravitational
multiplet, are the internal and external components
of the full $G^{(r)}$ gauge fields and the internal and external components 
of the $U (1)^{(16-r)}$ gauge fields. 
According to the argument given in  section \ref{ssft}, 
the  effective field theory 
of such fields will have a global $O(d,d+16-r;\mathbb{R})$ duality
symmetry to all orders in $\alpha'$.  
 The two-derivative version 
of this action is given by the same expression 
(\ref{hetACTIONFINAL}), with $\HH$ now
interpreted as a symmetric $O(d,d+16-r)$ matrix and $\CC$ interpreted as
a $(2d+16-r) \times \hbox{dim}\, G^{(r)}$ matrix, transforming as a vector of
$O(d, d+16-r)$ and as an adjoint of $G^{(r)}$. 

This paper is organized as follows.  In section 2 we give the
string theoretic arguments for the global duality symmetries 
of the effective field theories of heterotic massless fields.
The power of this argument is that it 
works to all orders in $\alpha'$.  We turn in section 3
to the torus compactification of the heterotic supergravity action,
including the
effect of the non-abelian gauge fields.  In section 4 we recast this 
action in terms of a formal $O(d, d+K)$ symmetry, with $K$ the dimension
of the gauge group.  Section 5 gives a rewriting of this theory in terms
of $O(d,d)$ multiplets, making this symmetry manifest.  In 
section 6 we discuss the possible relevance of our analysis for double
field theory formulations of heterotic strings that include $\alpha'$
corrections.

 \section{String theoretic argument} \label{ssft}
 
 We shall review the string theoretic argument for the existence of 
 $O(d,d)$
 symmetry in the presence of non-abelian gauge fields\cite{9109038}. 
 This argument is valid
 in classical string theory to all orders in $\alpha'$. 
  The main idea is to determine the symmetries of the action in a consistently
 truncated sector
 by studying the symmetries of the S-matrix in the same sector. 
 We shall then combine this with the obvious symmetries of the effective 
 action
 -- the $GL(d)$ symmetry associated with the linear transformation of the compact
 coordinates and the shift symmetry of the 2-form field -- to determine the
 full symmetry group of the truncated effective action. 
 The latter symmetries are not visible as symmetries of the
 S-matrix since they are typically spontaneously broken in a given background. 
 
 The theory
 under consideration is heterotic string theory and the truncation we
 are interested in requires 
 all fields to be  independent of $d$ of the spatial coordinates.
 The corresponding S-matrix will involve external states which carry zero momentum
 along the $d$ directions but has no further restrictions. Also since we shall be 
 interested in the classical effective action where we have integrated out all the massive
 string fields, it is sufficient to examine the S-matrix with massless external states only.

 While we shall
 consider a general set of external states subject to 
 the condition of independence of the $d$ spatial coordinates,
 we shall work in a 
 special background   left invariant by a large subset of the duality
  symmetries:  the two-form field and all gauge fields are zero and the
 metric is the diagonal unit metric.   
Working with such special background may seem a strong assumption 
but it is not so.   
 Once we have determined the symmetries of the S-matrix and translated 
 them into a symmetry statement for the effective action around the special background, the symmetry must also hold for the effective action in the more general backgrounds that can be obtained by switching on fields within the truncated class.  This is true even if the symmetry is spontaneously
 broken in the new background and is therefore not a symmetry
 of the S-matrix.  In our case, since the massless set of states include
 those for the internal components of the metric, two-form, and
 non abelian gauge fields, the general backgrounds for which these
 have expectation values are covered in the argument. 

Let us denote by $X^\mu$ the space-time coordinates on which the fields are allowed
to depend and by $Y^m$
the $d$ coordinates on which the fields do not depend. 
We also denote by $\psi^\mu$ and $\chi^m$ their fermionic partners. 
The vertex operators of the massless bosonic states  
in the minus-one picture  are then given by
\begin{equation} \label{evertex}
\begin{split}
& \psi^\mu \,\bar \p X^\nu\,  e^{ik\cdot X}e^{-\phi}, \\
& \psi^\mu\, \bar \p Y^n\,  e^{ik\cdot X}e^{-\phi},  \quad 
\chi^m \,\bar \p X^\mu e^{ik\cdot X} e^{-\phi}, \\
& \chi^m \,\bar \p Y^n \,e^{ik\cdot X}e^{-\phi},  \\
& \psi^\mu \, \bar J^\alpha \, e^{ik\cdot X}e^{-\phi}\  , \quad \ \ 
\chi^m \, \bar J^\alpha \, e^{ik\cdot X}e^{-\phi}\, ,
\end{split}
\end{equation}
where $\phi$ is the scalar arising from bosonization of the superconformal ghost 
system\cite{FMS}
and the $\bar J^\alpha$ are dimension (1,0) vertex operators in the anti-holomorphic sector
describing the $E_8\times E_8$ or $SO(32)$ currents. 
The vertex operators on the first line include those for the 
lower-dimensional metric, two-form, and dilaton.  On the
second line we have those for the lower-dimensional
abelian gauge fields that
arise from the metric and the two-form.  On the third line
we have the vertex operators for the scalar fields that arise
from the internal components of the metric and two form.
On the last line we have the vertex operators for 
the lower-dimensional non-abelian
gauge fields (first term) and for the scalars arising from the
internal components of the non-abelian gauge fields. 
None of the vertex operators in the above list carry momentum
or winding along the $d$ compact coordinates.

 The S-matrix will be computed from
the correlation function of these vertex operators together with 
suitable insertions of  picture changing
operators. The (holomorphic) 
picture changing operator with picture number plus one has the form
\be
- e^{\phi} (\psi_\mu \p X^\mu + \chi_m \p Y^m) + \cdots\,, 
\ee
where $\cdots$ denotes terms involving only ghost sector fields.
In the following we shall focus specifically on the tree level S-matrix which requires correlation functions of the conformal field theory on the sphere.
Now the key observations are
the following:
\begin{enumerate} 
\item 
In computing sphere correlation functions of  operators 
 in \refb{evertex} and
picture changing operators, we can treat the $d$ internal coordinates associated with the fields $Y^m$ as if they were non-compact. 
The compactness of these coordinates will affect
the correlation functions of vertex operators carrying non-zero momentum or winding number along these directions 
as well as higher-genus correlation functions of the vertex
operators given in \refb{evertex},  but not the correlation functions of the vertex
operators in \refb{evertex} on the sphere. \label{i1}
\item The 
correlation functions of the vertex operators 
\refb{evertex}, 
picture changing operators,
 and the additional ghost insertions needed to provide  
the correct integration measure over the moduli space of the punctured sphere can be expressed as a sum of  
 correlators 
 each of which factorizes into three factors: a correlator 
 involving $(Y^m,\chi^m)$'s, a correlator involving the
 $\bar J^\alpha$'s, and a correlator involving the other conformal fields.
  
\item On the sphere the correlation functions of the $Y^n$'s
satisfy holomorphic factorization. 
As a result
a correlation function involving the $(\p Y^m,\bar \p Y^m)$'s 
further factorizes into a correlation
function involving $\p Y^m$ and one involving $\bar \p Y^n$. 
This allows us to express the correlation functions of vertex operators in
\refb{evertex}, picture changing operators and other ghost insertions as sum of
terms each of which has
four parts:  a correlator 
 involving $(\p Y^m,\chi^m)$'s, a correlator involving $\bar\p Y^m$'s, 
 a correlator involving the
 $\bar J^\alpha$'s, and a correlator involving the other conformal fields.

\item 
Given   point \ref{i1},  we can compute the
required correlators in a theory where $Y^m$'s are  non-compact. In this case both the world-sheet theory for the $(Y^m,\chi^m)$ fields 
and 
the picture
changing operator are $O(d)$ invariant, where $O(d)$ acts as simultaneous rotation 
of the $Y^m$'s and $\chi^m$'s. Thus the factor in the
correlator involving the $(\p Y^m,\chi^m)$'s 
is $O(d)$ invariant.  
Furthermore due to holomorphic factorization
the factor involving
the $\bar \p Y^n$'s must also have
an independent $O(d)$ symmetry. 
 As needed, this symmetry is
also a (trivial) symmetry of the picture changing
operator, which does not involve the operator $\bar \p Y^n$.
In summary, we have an $O(d)\times O(d)$ symmetry.   

\item
The vertex operators given in \refb{evertex} provide a
representation of this $O(d)\times O(d)$ symmetry, 
\i.e.\ the action of $O(d)\times O(d)$ does not
take us outside this list. 
Thus the correlation functions and hence the S-matrix elements 
must have $O(d)\times O(d)$ symmetry.

\item If we expand the tree level S-matrix  
elements of massless states in string theory 
in powers of $\alpha'$, then to any given order in $\alpha'$ we can find  
a general coordinate invariant effective action whose tree level S-matrix elements
coincide with those computed from string theory. 
The $O(d)\times O(d)$ symmetry of the S-matrix elements then 
implies that the effective action that reproduces this S-matrix must also
have $O(d)\times O(d)$ symmetry. Furthermore at the linearized level the action of this
symmetry on the massless fields can be read out from their action on the vertex operators.

\item The effective action that reproduces the tree level S-matrix 
elements of massless
fields in toroidally compactified string theory can be regarded as 
the restriction of a general covariant action in 9+1 dimensions
to field configurations independent of $d$ coordinates. 
Since a general linear transformation on the $d$ coordinates preserves the property 
that the field configuration is independent of these $d$ coordinates, it must be a symmetry of the
resulting action. More precisely, in a generic theory of this kind, $GL(d)$ is only a symmetry
of the equation of motion and its $SL(d)$ subgroup is a symmetry of the action since the
$\sqrt{\det G}$ factor in the Lagrangian density is not $GL(d)$ invariant. In tree level
string
theory, however, the change in $\sqrt{\det G}$ can be cancelled by a shift in the dilaton field making $GL(d)$
a symmetry of the effective action.

Note that not all of this $GL(d)$ symmetry preserves the background and
thus not all of it is a symmetry of the S-matrix.
The $O(d)$ subgroup of $GL(d)$ describing the rotation of the $y$-coordinates
preserves the background and is  a symmetry 
of the S-matrix.\footnote{Even this $O(d)$ 
symmetry is broken 
once we take into account the periodic
identification of the internal coordinates, 
but this effect is not visible at tree-level string
theory when we have an effective theory for states carrying zero 
momenta along the internal directions.}
 This can be identified as the diagonal subgroup of 
the $O(d)\times O(d)$ symmetry discussed above.  

\item The full symmetry of the effective action must include both $O(d)\times O(d)$ and
$GL(d)$. 
The diagonal subgroup of $O(d)\times O(d)$ lies within $GL(d)$; so the number
of independent generators we get this way is $d^2$ from $GL(d)$ and 
$d (d-1)/2$ from one of the $O(d)$'s. To this we must add the shift symmetry
of the two-form fields; these are also manifest symmetries of the dimensionally  
reduced action since in all terms in the effective action the two-form appears with an exterior derivative
acting on it. These are parametrized by $d\times d$ anti-symmetric matrices
and give $d(d-1)/2$ more independent generators. Together they account for the
$d(2d-1)$ independent generators of $O(d,d)$. 
Thus $O(d,d)$ must be a symmetry of the effective action
to all orders in the $\alpha'$ expansion.
\end{enumerate}

We would like to remark that 
instead of working with the S-matrix elements we could also work with any
string field theory of heterotic NS fields, such as~\cite{Berkovits:2004xh}. In that case our arguments
will directly imply the $O(d)\times O(d)$ symmetry of the string field theory action
when we restrict the 
string fields to carry zero momentum along $d$ of the spatial directions. Since the effective action 
is obtained from this 
by integrating out the massive string fields followed by possible field redefinitions, 
it will inherit the $O(d)\times O(d)$ symmetry. Combining this with $GL(d)$ and shift symmetries we
can prove the $O(d,d)$ symmetry of the effective action.

The S-matrix argument can be easily generalized 
to consider a truncation where we
allow only gauge fields inside a subgroup $G\times U(1)^p$ of 
$E_8\times E_8$ or $SO(32)$ to be switched on.   Let $\bar J^{\alpha'}$ 
denote the currents for $G$ and we represent the $p$ abelian currents
by $i\bar \partial {U}^k$, with $k=1, \ldots, p$, and $U^k$ new chiral 
world-sheet scalar fields.\footnote{Strictly speaking 
the fields $U^k$ do not exist as conformal fields but the currents 
$\bar\p U^k$ do, and all our manipulations will involve only the currents.} 
In this case the list of operators in (\ref{evertex}) is modified:
\begin{equation} \label{evertex9}
\begin{split}
& \psi^\mu \,\bar \p X^\nu\,  e^{ik\cdot X}e^{-\phi}, \\
& \psi^\mu\, \bar \p {\cal Y}^{\bar n}\,  e^{ik\cdot X}e^{-\phi},  \quad 
\chi^m \,\bar \p X^\mu e^{ik\cdot X} e^{-\phi}, \\
& \chi^m \,\bar \p {\cal Y}^{\bar n} \,e^{ik\cdot X}e^{-\phi},  \\
& \psi^\mu \, \bar J^{\alpha'} \, e^{ik\cdot X}e^{-\phi}\  , \quad \ \ 
\chi^m \, \bar J^{\alpha'} \, e^{ik\cdot X}e^{-\phi}\, ,
\end{split}
\end{equation}
with 
\be
\bar\partial {\cal Y}^{\bar n} \ = \ \{ \bar\partial Y^1, \ldots, \bar\partial Y^d, \bar\partial U^1 \,, \ldots \bar\partial U^p \}  \,, \quad  \bar n \ = \ 1, \ldots, d+p\,. 
\ee
The main effect has been to include the $U(1)$ currents into an extended
version $\bar \partial {\cal Y}$ of the $\bar\partial Y$ conformal fields. 
Having truncated the gauge group, we now have less massless states. 

We can now repeat the above arguments.  The correlation functions
factorize into correlators involving the $\bar J^{\alpha'}$ and the rest.  
As before, the correlators involving $(\p Y^m,\psi^m)$'s 
are $O(d)$ invariant.
Furthermore the correlators involving   
$\bar\partial {\cal Y}^{\bar n}$'s factor from the rest and 
have  an $O(d+p)$ symmetry. Thus the S-matrix
and the effective action has  $O(d)\times O(d+p)$ symmetry to all orders in
$\alpha'$. 
We need to combine this with the manifest $GL(d)$ symmetry, the shift symmetry
of the 2-form fields parametrized by $d\times d$ anti-symmetric matrices and the
shift symmetry of the internal components of the $p$ gauge fields parametrized
by $d \times p$ matrices.\footnote{Although the Chern-Simons terms are not
invariant under the shift symmetry of the gauge fields, the three form field
strength can be made invariant under this transformation by including 
a compensating transformation of the 2-form fields.}
Taking into account that the diagonal $O(d)$ of $O(d)\times O(d+p)$ is
included in $GL(d)$, we get altogether 
\be   
d^2 + \tfrac{1}{2} (d+p)(d+p-1) +\tfrac{1}{2}  d(d-1) +dp
= \tfrac{1}{2}(2d+p)(2d+p-1)
\ee 
generators,  which is the right number of generators of
$O(d,d+p)$.

\section{Torus compactifcation with non-abelian gauge groups}\setcounter{equation}{0}
We perform the torus compactification of the spacetime action of heterotic strings for the massless fields,  
to zeroth order in $\alpha'$, but including all non-abelian gauge fields for a group $G$. Our goal is  
to investigate which global duality symmetry emerges once the massive Kaluza-Klein modes 
are truncated out. In the first subsection we make some general remarks on torus compactification 
or dimensional reduction and 
the nature (or rather absence) of flux quantization conditions. Then we present the 
technical details of the Kaluza-Klein reduction, 
which will be used in the following sections in order to write the action 
in terms of $O(d,d+{\rm dim}\,G)$ and $O(d,d)$ multiplets, respectively.

\subsection{Remarks on flux quantization}
Dimensional reduction, as distinguished from compactification,  is generally understood as a procedure
in which a theory formulated in a $D$-dimensional space-time
is used to construct a $D-p$ dimensional field theory with 
$0< p < D$.  This is done
by assuming that all fields are independent of $p$ spatial dimensions
and evaluating the original action with this assumption.  The nature
of the extra dimensions is left unspecified and any volume of the 
extra dimensions is taken to be a constant that can be absorbed in the
normalization of the action, sometimes as a rescaling of a coupling constant.

In order for dimensional reduction to produce a theory that is 
physically related to the original higher-dimensional theory, one
must specify the shape of the extra dimensions;  one must do compactification.   The simplest
compact $p$-dimensional space in which fields can consistently
be set to be constant is the $p$-dimensional torus $T^p$.  Even this
is not completely obvious for the cause of gauge fields, as we will
discuss below.   The dimensionally reduced theory is then obtained from the compactified theory by ignoring all 
Kaluza-Klein excitations that arise 
from field configurations in which fields depend on the compact space.   
Thus we view dimensional reduction as compactification on tori.

When an abelian gauge theory is defined on a torus there are
configurations where the gauge fields are not constant over the torus
and as a result there are non-vanishing field strengths.  The total flux
associated with an abelian field strength is quantized because only then space-dependent gauge fields on the torus are well-defined globally.  

For compactification on a torus we will consider the ansatz in which
all higher-dimensional non-abelian 
gauge fields 
$ \hat A_{\hat \mu}{}^{ \alpha}$  
are independent of the
toroidal directions $y^m$.  Letting $a_m{}^\alpha$ denote
the  components of the non-abelian 
gauge fields along toroidal directions, the field strength ${F}_{{m}{n}}{}^{\alpha}$ 
along  toroidal directions is then given by
\be  
 {F}_{{m}{n}}{}^{\alpha} \ \ = \ \partial_{m}a_{n}{}^{\alpha} 
   -\partial_{n}a_{m}{}^{\alpha} +f^{\alpha}{}_{\beta\gamma}
   a_{m}{}^{\beta}
   a_{n}{}^{\gamma} \ = \ f^{\alpha}{}_{\beta\gamma}
   a_{m}{}^{\beta}
   a_{n}{}^{\gamma} \,.  
\ee
It is now clear
that the dimensional reduction hypothesis of coordinate independence can lead to non-vanishing non-abelian field strengths.  This could not happen for abelian gauge fields, where only spatial dependence can lead to 
field strengths.  Moreover, unless the fields $a_m{}^{ \alpha}$ satisfy unusual constraints, the associated fields strengths will actually take arbitrary continuous values.  We claim  that there is no condition on the constant non-abelian gauge fields on the torus, and no quantization of the resulting fluxes.  This is simply because constant gauge fields on a torus are globally well-defined regardless of their value: they require no gauge transformation to patch up as we traverse any non contractible closed loop on the torus.   This means that we can perform the dimensional reduction without topological complications. 

It should be noted that   
in general 
spatially varying non-abelian gauge field  
configurations may require a quantization condition to be globally well defined, resulting in quantized fluxes.  
Here 
we see that non-abelian field strengths arising from spatial derivatives are not on the same footing as field strengths arising from the commutator term (for which there is no quantization, if the connections are spatially constant).  Indeed, one can find a simple example of non-abelian $SU(2)$ gauge fields where gauge fields with spatial dependence and gauge field without spatial dependence give rise to the same field strength.  These configurations are not even locally gauge equivalent.

\subsection{Torus compactification of heterotic supergravity}
We now perform the explicit compactification starting from the heterotic spacetime 
action. Even though this theory is defined in 10 space-time dimensions we shall  
keep our analysis slightly more general by taking the initial space-time dimension to be $D$.
Denoting the $D$-dimensional objects and indices by hats, the action is given by 
 \be \label{eDdim}
  S \ = \ \int {\rm d}^{D}x\sqrt{-\hat{g}}\, 
  e^{-2\hat{\phi}}\, \Big[\hat{R}+4(\partial\hat{\phi})^2 
  -\tfrac{1}{12}\,\hat{H}^{\hat{\mu}\hat{\nu}\hat{\rho}}\hat{H}_{\hat{\mu}\hat{\nu}\hat{\rho}}
  -\tfrac{1}{4}\, \hat{F}^{\hat{\mu}\hat{\nu}\alpha}\hat{F}_{\hat{\mu}\hat{\nu}\alpha}\Big].
 \ee 
The Einstein-Hilbert and dilaton terms are unchanged compared to the abelian case, 
but the field strengths are now 
 \be
  \begin{split}
   \hat{H}_{\hat{\mu}\hat{\nu}\hat{\rho}} \ &= \ 3
   \Big(\, \partial_{[\hat{\mu}}\, \hat{b}_{\hat{\nu}\hat{\rho}]}
   -\hat{A}_{[\hat{\mu}}{}^{\alpha}\partial_{\hat{\nu}}\hat{A}_{\hat{\rho}]\,\alpha} 
   -\tfrac{1}{3}\, f_{\alpha\beta\gamma}\hat{A}_{\hat{\mu}}{}^{\alpha}\hat{A}_{\hat{\nu}}{}^{\beta}
   \hat{A}_{\hat{\rho}}{}^{\gamma}\Big)\;, \\
   \hat{F}_{\hat{\mu}\hat{\nu}}{}^{\alpha} \ &= \ \partial_{\hat{\mu}}\hat{A}_{\hat{\nu}}{}^{\alpha} 
   -\partial_{\hat{\nu}}\hat{A}_{\hat{\mu}}{}^{\alpha} 
   +f_{\beta\gamma}{}^{\alpha}  
      \hat{A}_{\hat{\mu}}{}^{\beta}
   \hat{A}_{\hat{\nu}}{}^{\gamma}\;, 
  \end{split}
 \ee 
where $\alpha,\beta$ are the adjoint indices of 
the Lie algebra associated with the gauge group. 
With Lie algebra generators $T_\alpha$ we have 
$[T_\alpha, T_\beta] = f_{\alpha\beta}{}^\gamma  T_\gamma$,
where $f_{\alpha\beta}{}^\gamma$ are the structure  
constants. For semisimple gauge algebras  we use the Cartan-Killing metric 
\be
\label{cartan-killing}
\kappa_{\alpha\beta}\ \equiv \  - \hbox{tr} \, (\hbox{ad} \, T_\alpha \circ \hbox{ad} \, T_\beta )  \ = \ -  f_{\alpha \gamma}{}^\delta \, f_{\beta \delta}{}^\gamma \,,
\ee to lower indices, leading to $f_{\alpha\beta\gamma} \equiv 
f_{\alpha\beta}{}^\rho \kappa_{\rho \gamma}$ that is totally antisymmetric.
The inverse of $\kappa_{\alpha\beta}$ exists and is written as 
$\kappa^{\alpha\beta}$.
Note also that the overall normalization of  $\kappa$ 
is not important as it can be absorbed into a rescaling of the metric, dilaton and two form fields.
If  the gauge group is of the form $G'\times U(1)^p$ with the 
Lie algebra of   
$G'$ semisimple, then $\kappa$ 
is defined to be  a block diagonal matrix containing  
the $p\times p$ identity matrix $I_p$ and the
Cartan-Killing metric  $\kappa'$ for $G'$: 
\be
\label{kappa-prime}
\hbox{For} \ \  G' \times U(1)^p :  \qquad 
\kappa \ \equiv \  \begin{pmatrix}  I_p & 0 \\ 0 & \kappa' \end{pmatrix}
\,.  \ee 
This $\kappa$  matrix is still  invertible.

We perform the 
dimensional reduction by splitting the coordinates into non-compact and compact ones, 
corresponding to a toroidal background  
 \be
  \mathbb{R}^{n-1,1}\times T^d\;, \qquad D=n+d\;. 
 \ee
Specifically, we write $x^{\hat{\mu}}=(x^{\mu},y^m)$, 
corresponding to the index split
 \be
  \hat{\mu} \ = \ (\mu, m)\;, \qquad \hat{a} \ = \ (a,\underline{a})\;, 
 \ee
where the second equation indicates 
the splitting of the flat (Lorentz) indices.   The Lorentz metric is 
\be
\hat \eta_{\hat a \hat b} \ = \ \begin{pmatrix} 
\eta_{ab} & 0  \\[0.3ex]  
0 &  \delta_{\underline{a} \, \underline{b}} \end{pmatrix}  \,,
\ee
 with $\eta_{ab}$ for the noncompact directions and
$\delta_{\underline{a}\, \underline{b}}$ for the compact ones.
The Kaluza-Klein ansatz for the vielbein  $\hat{e}_{\hat{\mu}}{}^{\hat{a}} $ 
(and its inverse $ \hat{e}_{\hat{a}}{}^{\hat{\mu}}$)  is   
\be\label{KKBEin}
  \hat{e}_{\hat{\mu}}{}^{\hat{a}} \ = \ \begin{pmatrix} e_{\mu}{}^{a} & A_{\mu}^{(1)m}E_{m}{}^{\underline{a}} \\ 
  0 & E_{m}{}^{\underline{a}} \end{pmatrix}\;, \qquad
   \hat{e}_{\hat{a}}{}^{\hat{\mu}} \ = \ \begin{pmatrix} e_{a}{}^{\mu} & -e_{a}{}^{\nu}A_{\nu}^{(1)m} \\ 
  0 & E_{\underline{a}}{}^{m} \end{pmatrix}\;,  
\ee
where  $e_\mu{}^a$ and $e_a{}^\mu$ are inverses of each other, 
$E_{m}{}^{\underline{a}}$ and $E_{\underline{a}}{}^{m}$ are inverses of each other, and $A_\mu^{(1)m}$ denote a collection
of  Kaluza-Klein vectors labelled by $m$.  We define 
\be
g_{\mu\nu} \ \equiv \  e_\mu{}^a e_\nu{}^b \eta_{ab} \,, \qquad
G_{mn} \ \equiv \  E_{m}{}^{\underline{a}}E_{n}{}^{\underline{b}}\,  \delta_{\underline{a}\, \underline{b}}\,.  
\ee
In terms of these we have, 
with $\hat{\mu}  =  (\mu, m)$ and $\hat{\nu}  =  (\nu, n)$,
\be
\hat g_{\hat \mu \hat \nu}  \ \equiv \  \hat e_{\hat\mu}{}^{\hat a} 
\hat e_{\hat\nu}{}^{\hat b}  \hat\eta_{\hat a \hat b} \ = \ 
\begin{pmatrix} g_{\mu\nu}+ A_\mu^{(1)m} G_{mn} \, A_\nu^{(1)n} \ 
 & A_{\mu}^{(1)k}G_{kn} \\[0.7ex] 
 A_{\nu}^{(1)k}G_{km} & G_{mn} \end{pmatrix}\;. 
\ee

  In order to obtain canonically normalized and manifestly gauge invariant kinetic terms in the reduced theory 
  we have to perform a number of field redefinitions for the vectors and two-forms. 
  The general prescription, as also employed by Maharana-Schwarz, 
  is to define components of the $D$-dimensional fields with flat indices
  and then to `un-flatten' with the lower-dimensional vielbein.  
  This is best explained
  using an object with a single index, as the generalization to
  multiple indices is trivial.   Given an object $\hat W_{\hat\mu}$ 
  with $\hat\mu = (\mu, m)$ we define 
  \be
  \label{index-ms}
  \begin{split}
  W_m \ \equiv \ &\,  \hat W_m\,, \\
  W_\mu  \ \equiv \  &\,   e_\mu{}^a  \, \hat e_a{}^{\hat \nu} \hat W_{\hat \nu} \ = \  e_\mu{}^a  \,  \hat W_{a}\,.  
  \end{split}
  \ee
  Using the explicit form of the vielbein we get
  \be
  W_\mu  =  e_\mu{}^a  \bigl( \hat e_a{}^\nu   \hat W_\nu  + 
 \hat  e_a{}^{n}  \,   \hat W_{n}  \bigr) \ = \ 
  \hat W_\mu \,  - \,  e_\mu{}^a e_a{}^\nu  A_\nu^{(1)n} \,\hat W_n\,, 
 \ee
 and therefore
 \be
 W_\mu  \ = \ 
  \hat W_\mu \,  - \,    A_\mu^{(1)n} \hat W_n\, . 
 \ee
 When we deal with multiple indices we apply the rule in (\ref{index-ms})
 to each of the indices.  The logic behind the rule is that one can 
 quickly verify that
\be
\begin{split}
W^\mu W_\mu \ & \equiv \  g^{\mu \nu } W_\mu W_\nu  \ = \  \eta^{ab} \hat W_a  \hat W_b   \ \equiv \  \hat W^a \hat W_a\;,  \\
W^m W_m \ & \equiv \  G^{m n } W_m W_n   \ = \  \delta^{\underline{a}\, \underline{b}} \hat W_{\underline{a}}   \hat W_{\underline{b}}   \;,  
\end{split}
\ee
and this leads to 
\be
\label{index-expand}
\hat W^{\hat \mu}  \hat W_{\hat \mu} \ \equiv \ \hat g^{\hat \mu\hat \nu}
\hat W_{\hat \mu} \hat W_{\hat \nu}  \ = \ \hat \eta^{\hat a\hat b}
\hat W_{\hat a} \hat W_{\hat b} \ = \ W^\mu W_\mu +  W^m W_m\,,
\ee 
 giving a very simple way to expand contracted full-dimensional indices, without 
 off-diagonal metric contributions involving bare Kaluza-Klein vectors.

We turn now to the decomposition of the gauge kinetic terms. 
For the field strength $\hat H_{\hat\mu\hat\nu\hat \rho}$ we have,
for example,  
 \be
  H_{\mu mn} \ \equiv \  {e}_{\mu}{}^{a}\hat{e}_{a}{}^{\hat{\nu}}\hat{H}_{\hat{\nu}mn}  \quad \to \quad H_{\mu mn} \ \equiv \ \hat{H}_{\mu mn}-A_{\mu}^{(1)k}\hat{H}_{kmn}\,. 
 \ee 
For the full set of components we find
 \be
  \begin{split}
  H_{mnk} \ & \equiv \ \hat{H}_{mnk}\,, \\
  H_{\mu mn} \ &  \equiv \ \hat{H}_{\mu mn}
  -A_{\mu}^{(1)k}\hat{H}_{kmn}\,, \\
   H_{\mu\nu m} \ &  \equiv  \ \hat{H}_{\mu\nu m}-2A_{[\mu}^{(1)n}\hat{H}^{}_{\nu]mn}
   +A_{\mu}^{(1)n} A_{\nu}^{(1)k} \hat{H}_{mnk}\;, \\
   H_{\mu\nu\rho} \ & \equiv \ \hat{H}_{\mu\nu\rho}-3A_{[\mu}^{(1)m}\hat{H}^{}_{\nu\rho]m}
   +3A_{[\mu}^{(1)m}A_{\nu}^{(1)n}\hat{H}^{}_{\rho]mn}-A_{\mu}^{(1)m}A_{\nu}^{(1)n} A_{\rho}^{(1)k}\hat{H}_{mnk}\;.  
  \end{split}
 \ee  
Analogous redefinitions are needed for the Yang-Mills field strength: 
 \be
 \begin{split}
{ F}_{mn}{}^{\alpha} \ & \equiv \  \hat{F}_{mn}{}^{\alpha}\, , \\
  { F}_{\mu m}{}^{\alpha} \ &\equiv \ \hat{F}_{\mu m}{}^{\alpha}+A_{\mu}^{(1)n}\hat{F}_{mn}{}^{\alpha}\;, \\
  { F}_{\mu\nu}{}^{\alpha} \ &\equiv \  \hat{F}_{\mu\nu}{}^{\alpha}+2A_{[\mu}^{(1)m} \hat{F}^{}_{\nu]m}{}^\alpha
   +A_{\mu}^{(1)m} A_{\nu}^{(1)n} \hat{F}_{mn}{}^\alpha \;. 
  \end{split}
 \ee  
 
Our formula (\ref{index-expand}) makes the expansion of 
kinetic terms trivial.   It follows that
the Yang-Mills kinetic term decomposes as
  \be\label{YMDECOM}
  -\tfrac{1}{4}\hat{F}^{\hat{\mu}\hat{\nu}\alpha}\hat{F}_{\hat{\mu}\hat{\nu}\alpha} \ = \ -\tfrac{1}{4}{ F}^{\mu\nu\alpha} { F}_{\mu\nu\alpha}-\tfrac{1}{2}{ F}^{\mu m\alpha} { F}_{\mu m\alpha}
  -\tfrac{1}{4}{ F}^{mn\alpha} { F}_{mn\alpha}\;. 
 \ee 
Similarly, for the two-form kinetic term:  
 \be\label{HsQdecom}
    -\tfrac{1}{12}\hat{H}^{\hat{\mu}\hat{\nu}\hat{\rho}}\hat{H}_{\hat{\mu}\hat{\nu}\hat{\rho}}
    \ = \ -\tfrac{1}{12}H^{\mu\nu\rho}H_{\mu\nu\rho}-\tfrac{1}{4}H^{\mu\nu m} H_{\mu\nu m}
    -\tfrac{1}{4}H^{\mu mn}H_{\mu mn}-\tfrac{1}{12}H^{mnk} H_{mnk}\;. 
 \ee   
The index contractions above are done using the metrics $g_{\mu\nu}$ 
and $G_{mn}$.  In the two equations above, the new terms compared to MS are 
those with purely internal coordinates: ${ F}_{mn\alpha}^2$  
and $H_{mnk}^2$.  These vanish when fields are
$y$ independent and the gauge group is abelian but are non-zero
when the gauge group is non-abelian.  These terms simply give the potential:
 \be\label{redPOT}
  -V \ = \ -\tfrac{1}{12}H^{mnk} H_{mnk}-\tfrac{1}{4}F^{mn\alpha} F_{mn\alpha}\;. 
 \ee

The little less trivial part of the computation is to express the above field strengths in terms of the gauge potentials that are redefined as well in order to exhibit the non-abelian 
symmetry in conventional form.   
For the gauge potentials the
original field $\hat A_{\hat \mu}{}^\alpha$ yield
fields $a_m{}^\alpha$ and $A_\mu{}^\alpha$  from the
postulated rule: 
 \be
   \begin{split}
   {a}_{m}{}^{\alpha}  \ &\equiv \ \hat{A}_{m}{}^{\alpha}\,, \\
     {A}_{\mu}{}^{\alpha} \ &\equiv \  \hat{A}_{\mu}{}^{\alpha} -A_{\mu}^{(1)m}\,
    \hat{A}_{m}{}^{\alpha}\;. 
   \end{split}
  \ee  
Solving for the hatted components we get
 \be
   \begin{split}
    \hat{A}_{m}{}^{\alpha} \ &= \ {a}_{m}{}^{\alpha}\;, \\
    \hat{A}_{\mu}{}^{\alpha} \ &= \ {A}_{\mu}{}^{\alpha}+A_{\mu}^{(1)m}\,
    {a}_{m}{}^{\alpha}\;. 
   \end{split}
  \ee  
From the two-form potentials $\hat b_{\hat\mu\hat \nu}$ we get
scalar fields $B_{mn}$,  lower-dimensional
abelian gauge fields $A_{\mu m}^{(2)},$ and a lower-dimensional
two-form $b_{\mu\nu}$ defined from the 
relations:\footnote{The last terms on the right hand side 
of the last two equations 
in  \refb{ebdefn99} 
are needed due to the presence of the Chern-Simons
term in the $D$ dimensional action. 
Also note that, apart from those terms,  the 
definition of $b_{\mu\nu}$ in terms of 
$\hat b_{\hat\mu\hat\nu}$  
differs from the prescription given earlier:   the second term on the right hand side 
has coefficient 1 rather than 2 and there is a missing  $A^{(1)} A^{(1)} \hat b$ term. 
As a result, $b_{\mu\nu}$ has an 
anomalous transformation under the gauge
symmetry associated with $A_\mu^{(1)m}$. However, it also has 
an anomalous transformation under the gauge symmetry associated with $A_\mu^{(2)}{}_m$, 
and only for the above redefinition do they combine into a manifestly $O(d,d)$ invariant
gauge symmetry.} 
\be \label{ebdefn99}
  \begin{split}
  {B}_{mn}  \ &\equiv \ \hat{b}_{mn}\;, \\
  \ A_{\mu m}^{(2)} \ &\equiv  \ \hat{b}_{\mu m} -A_{\mu}^{(1)k}\ \hat{b}_{km}+\tfrac{1}{2}\, a_{m}{}^{\alpha}A_{\mu\alpha}\;, \\
 {b}_{\mu\nu}   \ &\equiv \ \hat{b}_{\mu\nu} +A_{[\mu}^{(1)m}\,\hat b_{\nu]m}
   -\tfrac{1}{2} a_{m}{}^{\alpha}A_{[\mu}^{(1)m}A^{}_{\nu]\alpha}\;.
  \end{split}
 \ee  
Solving for the hatted components we find:
\be \label{ebdefn}
  \begin{split}
   \hat{b}_{mn} \ &= \ {B}_{mn}\;, \\
   \hat{b}_{\mu m} \ &= \ A_{\mu m}^{(2)}-A_{\mu}^{(1)n}{B}_{mn}-\tfrac{1}{2}\, a_{m}{}^{\alpha}A_{\mu\alpha}\;, \\
   \hat{b}_{\mu\nu} \ &=\ {b}_{\mu\nu}-A_{[\mu}^{(1)m}A_{\nu]m}^{(2)}
   +A_{\mu}^{(1)m} A_{\nu}^{(1)n} {B}_{mn}+a_{m}{}^{\alpha}A_{[\mu}^{(1)m}A^{}_{\nu]\alpha}\;.
  \end{split}
 \ee  
Note that the abelian gauge fields arising from the metric have the
internal index up while those arising from the antisymmetric two-form have the internal index down.  The superscripts $(1)$ and $(2)$ 
are thus not strictly needed, but they help 
distinguish those two sets. For these gauge fields we neither raise nor  
lower the internal index.  A straightforward but somewhat tedious computation gives for the field strength in terms of the
redefined fields, 
 \be
   \begin{split}
    H_{mnk} \ &= \ -f_{\alpha\beta\gamma}\, \, 
    a_{m}{}^{\alpha} a_{n}{}^{\beta} a_{k}{}^{\gamma}   \;,  \\
    H_{\mu mn} \ &= \ \partial_{\mu}B_{mn}+\, 
    a_{[m}{}^{\alpha}D_{\mu}a_{n]\alpha}\;, \\
    H_{\mu\nu m} \ &= \ {\cal F}_{\mu\nu m}^{(2)}-C_{mn} \,{\cal F}_{\mu\nu}^{(1)n}-a_{m}{}^{\alpha} {\cal F}_{\mu\nu\alpha}
    \;, 
   \end{split}
  \ee  
where  we use the covariant derivatives,  non-abelian field strengths,
abelian field strengths, and the auxiliary scalars  defined by   
 \be\label{nonABEL}
 \begin{split}
  D_{\mu}a_{m\alpha} \ &\equiv  \ \partial_{\mu}a_{m\alpha}-A_{\mu}{}^{\gamma}f_{\gamma\alpha\beta}\, 
  a_{m}{}^{\beta}\;, \\   
  {\cal F}_{\mu\nu \alpha} 
  \ &\equiv  \ 2\partial_{[\mu}A_{\nu]\alpha} 
  +f^{}{}_{\alpha\beta\gamma}A_{\mu}{}^{\beta}
  A_{\nu}{}^{\gamma}\, , \\
 {\cal F}_{\mu\nu}^{(1)m}\ & \equiv \  \partial_\mu {A}_{\nu}^{(1)m} 
 - \partial_\nu {A}_{\mu}^{(1)m}\,, \\ 
  {\cal F}_{\mu\nu m}^{(2)}\ & \equiv \  \partial_\mu {A}_{\nu m}^{(2)} 
 - \partial_\nu {A}_{\mu m}^{(2)}\,, \\   
  C_{mn} \ &  \equiv \ B_{mn}+\tfrac{1}{2}\, a_{m}{}^{\alpha} \,
  a_{n\alpha}\;. \end{split}
 \ee  
The most laborious part of the calculation is to verify that 
 \be \label{edefhmnr}
  H_{\mu\nu\rho} \ = \ 3\Big(\partial_{[\mu}\, 
  b_{\nu\rho]}-A_{[\mu}^{(1)m}\partial^{}_{\nu}A^{(2)}_{\rho] m}  
  -\partial^{}_{[\nu}A_{\rho}^{(1)m}  A_{\mu] m}^{(2)}  
     -{A}_{[{\mu}}{}^{\alpha}\partial_{{\nu}}{A}_{{\rho}]\,\alpha} 
   -\tfrac{1}{3}f_{\alpha\beta\gamma}{A}_{{\mu}}{}^{\alpha}{A}_{{\nu}}{}^{\beta}
    {A}_{{\rho}}{}^{\gamma}\Big)\,,
 \ee 
so that it can be combined into the $O(d,d)$ covariant form,  
to be given  in  (\ref{Recall3form}) below. 
Similarly, the components of the (redefined) Yang-Mills field strength become  
 \be
 \begin{split} 
  { F}_{\mu\nu}{}^{\alpha}  \ &= \ {\cal F}_{\mu\nu}{}^{\alpha}+
  {\cal F}_{\mu\nu}^{(1)m} \, a_{m}{}^{\alpha}\;, \\
  { F}_{\mu m}{}^{\alpha} \ &= \ D_{\mu}\, a_{m}{}^{\alpha}\;, \\
  { F}_{mn}{}^{\alpha} \ &= \ f^{\alpha}{}_{\beta\gamma} \,
  a_{m}{}^{\beta} a_{n}{}^{\gamma}\;. 
 \end{split}
 \ee
This is close to the Maharana-Schwarz result but there are some
differences. First,
all partial derivatives become covariant derivatives 
when acting on objects with $\alpha$ index. Second,
abelian field strengths 
become non-abelian 
field strengths. Finally, we
have  additional 
terms involving purely internal field strength components.

Given (\ref{redPOT}) and the above results, 
the last contribution can be expressed as 
 \be\label{redPOTs}
  -V \ = \ -\tfrac{1}{12}\, f_{\alpha\beta\gamma} f_{\alpha'\beta'\gamma'}\, a_{m}{}^{\alpha} a_{n}{}^{\beta} a_{k}{}^{\gamma}
 \, a^{m\alpha'} a^{n\beta'} a^{k\gamma'}-\tfrac{1}{4} 
 f^{\alpha}{}_{\beta\gamma} f_{\alpha\beta'\gamma'} 
 a_{m}{}^{\beta} a_{n}{}^{\gamma}  a^{m\beta'} a^{n\gamma'}\;. 
 \ee    
We summarize this section by assembling the pieces and giving the final form 
of the dimensionally reduced action:
\be
\label{ms-nonabelian}
 \begin{split}
  S \ = \ \int d^n x\sqrt{-g}\,e^{-2\phi}\Big[&\, R(g) +4
  \,\partial^\mu\phi \, \partial_\mu \phi 
  +\tfrac{1}{4}\, \partial^{\mu}G^{mn}\partial_{\mu}G_{mn}
  -\tfrac{1}{4}\, G_{mn}{\cal F}^{\mu\nu(1)m} {\cal F}_{\mu\nu}^{(1)n}\\
  &-\tfrac{1}{4}\,\kappa_{\alpha\beta}F^{\mu\nu\alpha} F_{\mu\nu}{}^{\beta}
  -\tfrac{1}{2}\kappa_{\alpha\beta} G_{mn}F^{\mu m\alpha} F_{\mu}{}^{ n\beta}\\
  &-\tfrac{1}{12}H^{\mu\nu\rho} H_{\mu\nu\rho}
  -\tfrac{1}{4}\, H^{\mu\nu m} H_{\mu\nu m}
  -\tfrac{1}{4} \, H^{\mu mn} H_{\mu mn} - V \Big]\;. 
 \end{split}
\ee  
Here the volume element takes the `string frame' canonical form thanks to the 
redefinition
 \be
  \phi \  = \ \wh{\phi}-\tfrac{1}{4}\log \det 
  G_{mn}\;. 
 \ee 
The terms in the first line of the action originate from the Einstein-Hilbert and dilaton terms in 
$D$ dimensions, which are not affected by the non-abelian gauge couplings and therefore 
can be taken directly from \cite{Maharana:1992my}. The terms in the second line 
originate from the Yang-Mills kinetic term, c.f.~(\ref{YMDECOM}), while the terms in 
the third line other than $V$  
originate from the kinetic term of the $b$-field in (\ref{HsQdecom}). 
Finally, the potential $V$ is given in (\ref{redPOTs}) and encodes the terms 
not present in the Maharana-Schwarz analysis  
(beyond those originating from covariantizing gauge couplings). 

In the next section we write the above 
 action in an  $O(d,d+K)$ covariant form, 
with $K={\rm dim}\,G$, although the theory only has $O(d,d)$ as a proper symmetry. 
For this purpose  we assemble the terms in the
above action in slightly different order  
\be
\label{ms-nonabelians}
 \begin{split}
  S \, = \, \int d^n x\sqrt{-g}\,e^{-2\phi}&\Big[\, R(g) +4
  \partial^\mu\phi \partial_\mu \phi -\tfrac{1}{12}H^{\mu\nu\rho} H_{\mu\nu\rho}
 \\ &
  +\tfrac{1}{4}\, \partial^{\mu}G^{mn}\partial_{\mu}G_{mn}
   -\tfrac{1}{2}\kappa_{\alpha\beta} G_{mn}F^{\mu m\alpha} F_{\mu}{}^{ n\beta}
    -\tfrac{1}{4} \, H^{\mu mn} H_{\mu mn}  \\[0.5ex]
    &
  -\tfrac{1}{4}\, G_{mn}{\cal F}^{\mu\nu(1)m} {\cal F}_{\mu\nu}^{(1)n}
  \ -\tfrac{1}{4}\,\kappa_{\alpha\beta}F^{\mu\nu\alpha} F_{\mu\nu}{}^{\beta} -\tfrac{1}{4}\, H^{\mu\nu m} H_{\mu\nu m}
- V  \Big]\;. 
 \end{split}
\ee

\section{Compactified theory in terms of $O(d,d+K)$ multiplets}
We now rewrite
the action \refb{ms-nonabelians} in a form that is covariant under $O(d,d+K)$, where $K$ is the {\em dimension} of the gauge algebra. 
The gauge algebra type will be discussed below.   
We will use the convention that indices   
and objects  
transforming covariantly 
under $O(d,d+K)$ are hatted. 
This should not be confused
with the use of hats in the previous section, where they refer to higher-dimensional objects and indices.  Furthermore 
from this section onwards we shall use the symbol $\wh\eta$ and $\eta$ to describe
respectively the $O(d,d+K)$ and $O(d,d)$ invariant metric and not the Minkowski
metric as in the last section.
We now claim that  the dimensionally reduced action \refb{ms-nonabelians} 
can be 
written as\footnote{This action is of the same structural form as that obtained 
by Scherk-Schwarz compactification of heterotic supergravity truncated to the Cartan subalgebra \cite{Kaloper:1999yr}.  
Moreover, it is closely related to that given in \cite{Lu:2006ah}, which considers group manifold reductions of heterotic supergravity including non-abelian gauge fields and also displays the action with a formal $O(d,d+K)$ symmetry.}  
 \be\label{hetACTION}
  \begin{split}
   S \ = \ \int d^nx\, \sqrt{-g}
      \,e^{-2\phi}\Big(&R(g)+4\, \partial^{\mu}\phi\, \partial_{\mu}\phi
   -\tfrac{1}{12}\, H^{\mu\nu\rho}H_{\mu\nu\rho} \\
   &+\tfrac{1}{8}\, D^{\mu}\wh {\cal H}^{\hat{M}\hat{N}}\,
   D_{\mu}\wh {\cal H}_{\hat{M}\hat{N}}
   -\tfrac{1}{4}\, \wh{\cal H}_{\hat{M}\hat{N}}\, 
   \wh {\cal F}^{\mu\nu \hat{M}}
   \wh {\cal F}_{\mu\nu}{}^{\hat{N}}-V(\wh {\cal H}) \Big)\;, 
  \end{split}
 \ee  
where   
\be\label{covDER}
  \begin{split}
   D_{\mu}\wh {\cal H}^{\hat{M}\hat{N}} \ &\equiv \ \partial_{\mu}\wh {\cal H}^{\hat{M}\hat{N}}-2\, \wh A_{\mu}{}^{\hat K}
   f_{\hat K}{}^{(\hat M}{}_{\hat L} \wh {\cal H}^{\hat N)\hat{L}}\;, \\
 \wh  {\cal F}_{\mu\nu}{}^{\hat M}  \ &\equiv \   \partial_{\mu}\wh A_{\nu}{}^{\hat M}-\partial_{\nu}\wh A_{\mu}{}^{\hat M}
   +f^{\hat M}{}_{\hat{K}\hat{L}} \wh A_{\mu}{}^{\hat{K}} \wh A_{\nu}{}^{\hat L}\;, \\
   H_{\mu\nu\rho} \ &\equiv \ 3\Big(\partial_{[\mu}b_{\nu\rho]}-
   \wh A_{[\mu}{}^{\hat M}\partial_{\nu}\wh A_{\rho]\hat M}
   -\tfrac{1}{3}\, f_{\hat{M}\hat{K}\hat{L}}\wh A_{[\mu}{}^{\hat M} 
   \wh A_{\nu}{}^{\hat K} \wh A_{\rho]}{}^{\hat L}\Big)\;, 
  \end{split}
 \ee  
and the potential is 
 \be\label{covPOTential}
  V({\cal H}) \  
  = \ \tfrac{1}{12}\, f^{\hat M}{}_{\hat K\hat P} f^{\hat N}{}_{\hat L\hat Q} \wh{\cal H}_{\hat M \hat N} \wh {\cal H}^{\hat K\hat L}
  \wh{\cal H}^{\hat P\hat Q}
  \, +\, \tfrac{1}{4}\, f^{\hat M}{}_{\hat N\hat K} f^{\hat N}{}_{\hat M\hat L} \wh {\cal H}^{\hat K\hat L}
  +\, \tfrac{1}{6}\, f^{\hat M\hat N\hat K} f_{\hat M\hat N\hat K}\;. 
 \ee 
Here the  $f^{\hat M}{}_{\hat N \hat K}$ 
are a set of constants which we shall call structure constants, 
and the indices  take values $\hat{M},\hat N,\ldots =1,\ldots,2d+K$.  
These indices are lowered and raised 
with a metric
$\wh\eta_{\hat M\hat N}$ 
 and its inverse $\wh\eta^{\hat M\hat N}\equiv (\wh\eta^{-1})^{\hat M\hat N}$:
\be\label{neweta}
\widehat\eta_{\hat M\hat N}  \ \equiv \ 
\begin{pmatrix} 0 & \delta^i{}_j & 0 \\ \delta_i{}^{j} & 0 &0 \\ 0 & 0 & \kappa_{\alpha\beta} \end{pmatrix}\;.
\ee
Since $\kappa$ is invertible, $\wh \eta$ is also invertible.  
Associated with the constant invertible metric $\wh \eta$ 
there is a set of matrices $\Omega$ that preserve it.  
With $\Omega$ carrying index structure $\Omega_{\hat M}{}^{\hat N}$
the matrices satisfying 
\be \label{eoddtrs}
\Omega \ \wh\eta \ \Omega^T \ = \  {\wh\eta}\, , 
\ee
form a group under multiplication. 
Because all indices are properly contracted, the action is
invariant under duality transformations 
\be
\begin{split}
\wh {\cal H}_{\hat M\hat N} \ \to  \  
(\,\Omega  \,\wh {\cal H} & \,\Omega^T\, )_{\hat M\hat N}\ , \quad   \ \ 
\wh A_\mu{}^{\hat M}
\ \to \  (\wh\eta^{\, -1} \Omega\,\wh \eta\, )^{\hat M}{}_{\hat N} \, 
\wh A_\mu{}^{\hat N}\, \,, \\
&    f_{\hat M\hat N\hat K } \ \to \    
\Omega_{\hat M}{}^{\hat M'}\Omega_{\hat N}{}^{\hat N'}
\Omega_{\hat K}{}^{\hat K'} f_{\hat M'\hat N'\hat K' } \,.
\end{split}
\ee
The action \refb{hetACTION} is 
invariant if we set the
structure constants to zero, but non-zero values of the
structure constants $f$  will typically break this to a subgroup.  

Let us now discuss the duality group that arises for the metric
$\wh \eta$ in (\ref{neweta}).  If the matrices $\Omega$ satisfying  \refb{eoddtrs}
are changed to $A^{-1} \Omega A$, with $A$ invertible, they still form the same group. This time, however, the invariant metric 
is changed $\wh \eta \to A \,\wh \eta 
\, A^T$.  If the Lie algebra of the theory is compact and semisimple the Cartan-Killing metric $\kappa$  is positive definite and there
is a matrix $\omega$ such that $\omega \,\kappa \, \omega^T = I_K$, 
with $I_K$ the $K\times K$ identity matrix.  It then follows that by
taking $A$ to be of the block-diagonal form $( I_d, \omega)$, the metric
$\wh \eta$ can be put in the form $(\eta , I_K)$, with $\eta$ the $O(d,d)$ metric.  We then recognize that for compact semisimple Lie algebras we
have the duality group $O(d, d+K)$.   The case when the gauge group
contains $U(1)$ factors will be discussed at the end of the section.
These are the situations we have in
mind, and we will simply speak of $O(d, d+K)$ as the duality group. 
We have introduced explicitly the Cartan-Killing metric, however, to
allow for the possibility of future generalizations, including non-compact
semisimple algebras, as we discuss in the
conclusions.

 In order to make contact with \refb{ms-nonabelians} we need the explicit
expressions for 
$\wh{\cal H}_{\hat M\hat N}$ and 
$\wh A_\mu{}^{\hat M}$ in terms of the fields obtained in the previous section after
dimensional reduction, and also the values of the structure constants
$f_{\hat M\hat N\hat K}$.
The matrix $\wh{\cal H}$ is parameterized by the 
internal (scalar) components $G, B$, and $a$ of the metric, $b$-field, 
and gauge fields, respectively as follows:
\be\label{newH}
\begin{split}
\wh {\cal H}_{\hat M\hat N} \  = \ & \    
\begin{pmatrix}   
 \wh {\cal H}^{mn} &  \wh{\cal H}^{m}{}_{n} & 
 \wh{\cal H}^{m}{}_{ \beta} \\[0.3ex] 
 \wh {\cal H}_{m}{}^n & \wh{\cal H}_{mn} & \wh{\cal H}_{m \beta} 
  \\[0.3ex]
   \wh{\cal H}_{\alpha}{}^{n}   
  & \wh{\cal H}_{\alpha n} & \wh{\cal H}_{\alpha \beta}
\end{pmatrix} \\
\  = \ & \  \begin{pmatrix} G^{mn} & - G^{mk} C_{kn}  & - G^{mk} a_{k \beta} \\[0.5ex]
- G^{nk} C_{km}  & \ \  G_{mn} + C_{km} G^{kl} C_{ln} + a_{m}{}^{\gamma} 
a_{n \gamma} \  \ & C_{km} G^{kl} a_{l \beta}
 + a_{m \beta}  \\[0.5ex]
 - G^{nk} a_{k \alpha} & C_{kn} G^{kl} a_{l \alpha} + a_{ n \alpha} & \kappa_{\alpha \beta}
 + a_{k \alpha} G^{kl} a_{l \beta}
\end{pmatrix} \;, 
\end{split}
\ee 
with
\be
C_{mn} = B_{mn} + \tfrac{1}{2}\, a_{m}{}^{\alpha} a_{n}{}_{\alpha}\,.
\ee 
It is easy to see that
the generalized metric $\wh{\cal H}$ with matrix element $\wh{\cal H}_{\hat M\hat N}$
satisfies:
 \be
 \label{group-condition}
  \wh {\cal H} \, \wh\eta^{-1} \ \in  \ O(d,d+K) 
\ \   \Longleftrightarrow \   \ \wh{\cal H}_{\hat M}{}^{\hat P}
  \wh{\cal H}_{\hat N}{}^{\hat Q} \, \wh{\eta}_{\hat P \hat Q}\ = \ 
  \wh{\eta}_{\hat M \hat N}\,.  
 \ee
The gauge fields $A_{\mu m}^{(2)}\, , A_{\mu}^{(1)m}$, and $A_{\mu}{}^{\alpha}$ of the previous section are combined into an $O(d,d+K)$ vector as
 \be
 \label{gfdef78}
 \wh  A_{\mu}{}^{\hat{M}} \ \equiv \ \big(\,A_{\mu m}^{(2)}\,,\; A_{\mu}^{(1)m}\,,\;A_{\mu}{}^{\alpha}\,\big)\;, 
 \ee 
and so are the corresponding field strengths, 
 \be
 \label{fsdef67}
 \wh  {\cal F}_{\mu\nu}{}^{\hat{M}} \ \equiv \ \big(\,{\cal F}_{\mu\nu m}^{(2)}\,,\; {\cal F}_{\mu\nu}^{(1)m}\,,\;{\cal F}_{\mu\nu}{}^{\alpha}\,\big)\;. 
 \ee 
 The first two field strengths are abelian while the final one takes the non-abelian form (\ref{nonABEL}). 
 Finally the structure constants are chosen to be
   \be\label{fform}
   f^{\hat M}{}_{\hat N\hat K} \ = \  \left\{
  \begin{array}{l l}
    f^{\alpha}{}_{\beta\gamma} & \quad \text{if\; $(\hat M,\hat N,\hat K)=(\alpha,\beta,\gamma)$}\\[0.3ex]
    0 & \quad \text{otherwise}\\     \end{array} \right. \;, 
\ee
with $\alpha,\beta$ denoting the $K$ gauge algebra directions and 
where $f^{\alpha}{}_{\beta\gamma}$
are the structure constants of the gauge group $G$.

We shall now show that the action \refb{hetACTION} indeed  
coincides with the dimensionally reduced action \refb{ms-nonabelians}. 
The first line on each of the two actions is exactly the same.  
The second line on \refb{ms-nonabelians} reproduces the 
 $\tfrac{1}{8} D\wh{\cal H}\, D\wh{\cal H}$ term in  \refb{hetACTION}.
Similarly, the third line on \refb{ms-nonabelians}, except for the potential $V$,  reproduces the 
 $-\tfrac{1}{4} \wh{\cal H}\, \wh{\cal F}
 \wh {\cal F}$ term in  \refb{hetACTION}.
The only difference so far
with the Maharana-Schwarz analysis is the 
presence of covariant derivatives and non-abelian field strengths instead of partial derivatives 
and abelian field strengths.  Since $f^{\hat M}{}_{\hat N\hat K}$
is non-trivial only in the gauge algebra directions it reproduces the 
non-abelian gauge structures of the reduced theory.  Finally, it is a straightforward computation 
to verify that the potential (\ref{covPOTential}) reproduces the potential (\ref{redPOTs}) 
of the reduced theory.   Inserting  (\ref{fform})  
into (\ref{covPOTential}) we have 
 \be
  V \ = \ \tfrac{1}{12}\,
  f_{\alpha\gamma\delta}\, 
  f_{\beta\epsilon\kappa} \, 
  \wh{\cal H}^{\alpha\beta}  
  \wh{\cal H}^{\gamma\epsilon} \wh{\cal H}^{\delta\kappa}
  +\tfrac{1}{4}\, f^{\alpha}{}_{\beta\gamma} f^{\beta}{}_{\alpha\delta}
  \wh{\cal H}^{\gamma\delta}+
  \tfrac{1}{6}f^{\alpha\beta\gamma}f_{\alpha\beta\gamma}\;. 
 \ee 
This can be simplified using the value of $\wh{\cal H}_{\alpha\beta}$
from (\ref{newH}), and the result is indeed (\ref{redPOTs}).
We finally note 
that in the limit $f_{\alpha\beta\gamma}\rightarrow 0$ the action reduces to 
that found by Maharana-Schwarz, in which case the theory is properly invariant under 
a global $O(d,d+K)$ symmetry.

If we are willing to accept the Maharana-Schwarz action as a valid starting point,
we could arrive at the action \refb{hetACTION} using the following short argument.
First of all we note that the $D$ dimensional action has terms quadratic in $f_{\alpha\beta\gamma}$,
linear in $f_{\alpha\beta\gamma}$ and independent of $f_{\alpha\beta\gamma}$. 
The MS action corresponds to terms independent of $f_{\alpha\beta\gamma}$, and as pointed
out above, the $f_{\alpha\beta\gamma}$ independent part of the action \refb{hetACTION} 
coincides with the MS action. Thus we only need to verify that the terms linear and
quadratic in $f_{\alpha\beta\gamma}$ are correct. Now by examining the $D$ dimensional
action \refb{eDdim} we see that all the terms linear in $f_{\alpha\beta\gamma}$ have  a single
derivative and all the terms quadratic in $f_{\alpha\beta\gamma}$ have no derivatives. Thus
the dimensionally reduced action must also have this property. We see that \refb{hetACTION}
does share this property. Thus if \refb{hetACTION} is not the correct dimensionally reduced
action then any additional term must share this property. Furthermore since both the
original action \refb{eDdim} and the dimensionally reduced action \refb{hetACTION} are
gauge invariant, any additional term must also be gauge invariant. It is easy to see that
it is impossible to write down a gauge invariant term with a single derivative involving the
fields which appear in \refb{hetACTION} or equivalently in \refb{ms-nonabelians}. This shows that
there are no additional terms with a single power of $f_{\alpha\beta\gamma}$. This leaves us
to check that \refb{hetACTION} reproduces correctly the derivative free 
terms quadratic in $f_{\alpha\beta\gamma}$, i.e.\ that the potential term \refb{covPOTential} 
is correct.\footnote{The form of the potential can also be read off from the $f$-dependent terms in the  
heterotic double field theory action given in \cite{Hohm:2011ex}.} 
As discussed earlier, this term comes from \refb{redPOT} and can be easily computed,
leading to \refb{covPOTential}. This shows that the action \refb{hetACTION} is the correct
dimensionally reduced action.

The action \refb{hetACTION} given at the beginning of this section applies with some modifications
when the gauge group is $G'\times U(1)^p$, with $G'$ semisimple.
As explained before, the $\kappa$ matrix then takes 
the block-diagonal form in  \refb{kappa-prime}, 
and the $\wh \eta$ metric in \refb{neweta}  now becomes 
\be\label{newetaG'}  
\hbox{For} \ \  G' \times U(1)^p :  \qquad\widehat\eta  \ \equiv \ 
\begin{pmatrix} 0 & I_d & 0 & 0  \\ I_d & 0 &0 & 0 
\\ 0 & 0 &I_p & 0  \\ 0 & 0 &  0 &  \kappa' \end{pmatrix}\;.
\ee
With $G'$ compact semisimple,  this
metric is associated with the duality group  $O(d, d+ p+ K')$, where
$K'$ is the dimension of $G'$.  The indices now run as
$\hat M, \hat N \, \ldots = 1, \ldots, 2d+ p+ K'$. 
   The gauge fields are now an $O(d,d+p+K')$ vector:
 \be
 \label{gfdef78aa} 
 \wh  A_{\mu}{}^{\hat{M}} \ \equiv \ \big(\,A_{\mu m}^{(2)}\,,\; A_{\mu}^{(1)m}\,,\; A_\mu{}^i \,, \; A_{\mu}{}^{\alpha'}\,\big)\;, 
 \ee 
where $A_\mu{}^i$, with $i = 1, \ldots , p$, are $p$ abelian gauge fields,
and $\alpha' = 1, \ldots , K'$.   The structure constants
$f^{\hat M}{}_{\hat N\hat K}$ vanish except when all indices take 
values on the $K'$ components associated with the Lie algebra of $G'$.
This time $\wh{\cal H}$ is a $(2d+ p+ K') \times (2d+ p+ K')$ matrix and
 \be
 \label{newhmatrix}
  \wh {\cal H} \ \wh\eta^{\,\,  -1} \ \in  \ O(\, d\,,\, d+p+ K')   \,.
 \ee
The parameterization of $\wh{\cal H}$ can be obtained from that
in \refb{newH} by letting the Lie algebra gauge indices run over two kinds of  values:
$\, \alpha = (i, \alpha')$, again,  with $i= 1, \ldots, p$,  and $\alpha' = 1, \ldots, K'$. 
Moreover, we take $\kappa_{ij} = \delta_{ij}$,  $\kappa_{\alpha' i} =
\kappa_{ i\alpha'}= 0$,
and $\kappa_{\alpha'\beta'}$ the matrix elements of $\kappa'$.

We have emphasized that $O(d, d+ K)$ (or  $O(d,d+ p + K')$)  
are formal duality symmetries of the reduced action.   
Let us now discuss, following \cite{Hohm:2011ex},  
 the surviving global duality symmetries
of the reduced action.   
Consider first the case where $G$ is compact semisimple and of dimension $K$.  We first note that in this case the tensor $f^{\hat M}{}_{\hat N\hat K}$  
in (\ref{fform}) 
is \textit{not}  $O(d,d+K)$ invariant. 
Since the tensor vanishes whenever an index takes any of the first $2d$ values, 
it is     
 invariant under the $O(d,d)$ subgroup that shuffles these directions
while leaving the gauge algebra directions inert.\footnote{More precisely,
the global subgroup leaving (\ref{fform}) invariant  is 
$O(d,d)\times G$,  with $G$ the rigid subgroup of the gauge group.}
Specifically, for the gauge groups relevant for heterotic string theory we have 
 \be
  G \ = \ { SO}(32)\quad {\rm or}  \quad    {E}_8\times {E}_8\quad \rightarrow \quad \text{global duality symmetry:}\;\;
  O(d,d)\;.
 \ee  
If the gauge group is of the form $G=G'\times U(1)^p$,   
with $G'$ semi-simple, 
the tensor $f^{\hat M}{}_{\hat N\hat K}$ 
 vanishes whenever an index takes any of the first  $2d+ p$
values.  
Consequently, it is invariant under the larger group $O(d,d+p)$, 
which is the true duality symmetry. 
For instance, if we truncate the heterotic theory gauge
group down to
$E_8\times {U}(1)^8$, the massless effective field theory on $T^d$
will have:
 \be
  G \ = \ { E}_8\times { U}(1)^8\quad \rightarrow \quad \text{global duality symmetry:}\;\;
  O(d,d+8)\;.
 \ee

\sectiono{Compactified theory in terms of $O(d,d)$ multiplets} \label{sodd}

In the previous sections we have considered
the heterotic string with its full non-abelian gauge group $G$
compactified on a torus.  The low-energy effective field theory
action was displayed with a formal 
$O(d,d+K)$ 
global
symmetry, with $K$ the dimension of the non-abelian gauge group.
We have also seen that the true global symmetry 
of the low energy 
effective action is $O(d,d)\times G$   
for
compactifications without Wilson lines, and
the gauge fields give rise to massless adjoint scalars and
lower-dimensional massless gauge fields of $G$. 
The purpose of this section is to make this symmetry manifest by
rewriting the
low-energy action (\ref{hetACTION}) in terms of proper $O(d,d)\times G$ multiplets, 
instead of the fictitious  
$O(d,d+K)$ multiplets.   
The fields that will be used are
\be
  \;\; {\cal H}_{MN}\, ,\quad {\cal C}_{M}{}^{\alpha}\;,  \quad
  \hbox{with constraints:} \quad  
  {\cal H}\, \eta\,  {\cal H} \ = \ \eta\,  , \quad (1+{\cal H}\,\eta) 
\, {\cal C} \ = \ 0\, .
 \ee

When some Wilson lines are included in the heterotic compactification, the gauge group $G$ can be broken to a group $G'\times U(1)^p$. 
The duality group of the low-energy effective theory for the 
massless fields is enhanced to
$O(d, d+p)$.  
The analysis of this section can also be generalized to 
make the $O(d,d+p)$
symmetry of the action manifest by using $O(d, d+p) \times G'$ multiplets.

\subsection{Introducing $O(d,d)$ field multiplets}

Let us consider the $(2d+K)\times (2d+K)$ 
 generalized metric of 
equation (\ref{newH}) written in block form as follows:
\be \label{epp1}
\wh{\cal H} \ = \ \begin{pmatrix} 
\wt {\cal H} & \wt {\cal C}\\[0.3ex]
 \wt {\cal C}^{\,T} & \wt {\cal N} \end{pmatrix}\,. 
\ee
With more explicit index notation
 \be \label{epp1explicit}
 \widehat {\cal H}_{\hat M \hat N} \ = \ \begin{pmatrix} \wt {\cal H}_{MN} & \wt {\cal C}_{M\beta}  \\[0.4ex]
  (\wt {\cal C}^{\,T})_{\alpha N}  & \wt {\cal N}_{\alpha\beta} \end{pmatrix} \, ,
 \ee
 where now the indices $M,N$ run over $2d$ values. Thus the matrix
 dimensions are as follows 
 \be
 \label{sizes} 
 \begin{split}
 \wh \eta \ ,\  \wh{\cal H} \ : \  & \  (2d+K) \times (2d+K)\,, \\
 \eta \ ,\ \wt{\cal H} \ : \  & \  (2d) \times (2d) \,,\\
 \wt{\cal C} \ : \  & \  (2d) \times K\,,   \\
\kappa\ , \  \wt{\cal N} \ : \  & \  K \times K  \,.  
 \end{split}
 \ee 
Since $\widehat {\cal H} \, \wh\eta^{\ -1}$ is 
an $O(d, d+K)$ matrix it satisfies $\wh {\cal H} \, \wh\eta^{\ -1} \wh{\cal H} = \wh \eta$ (see \refb{group-condition}) and therefore 
\be\label{epp2}
\begin{pmatrix}\wt {\cal H} & \wt {\cal C}\cr \wt {\cal C}^T & \wt {\cal N}
\end{pmatrix}  \begin{pmatrix}  \eta & 0\cr 0 & \kappa^{-1} \end{pmatrix}
\begin{pmatrix} \wt {\cal H} & \wt {\cal C} \cr \wt {\cal C}^{\,T} & \wt {\cal N}  \end{pmatrix}  
\ = \ \begin{pmatrix} \eta & 0\cr 0 & \kappa \end{pmatrix} \, ,
\ee
where 
\be 
\eta \ \equiv \
\begin{pmatrix}
0 & \delta^i{}_j \cr \delta_i{}^j & 0
\end{pmatrix}\;, 
\ee 
is the $O(d,d)$ invariant tensor. The equality \eqref{epp2} 
implies three conditions for the 
block matrices:
\be
\label{constraint-eqnss}
\begin{split}
\wt {\cal H}\, \eta\,  \wt {\cal H }+ \wt {\cal C }\,\kappa^{-1}\, \wt {\cal C}^{\,T} \ = \ & \ \eta\,,\\[0.5ex]
\wt {\cal H}\, \eta \, \wt {\cal C} + \wt {\cal C} \,\kappa^{-1}\, \wt {\cal N} \ = \ & \ 0 \,, \\[0.5ex]
\wt {\cal C}^{\,T}  \eta \, \wt {\cal C} + \wt {\cal N} \,\kappa^{-1} \,  \wt {\cal N} \ = \ & \ \kappa \,.
\end{split}
\ee

We 
shall now try to find a suitable parametrization of $\wt{\cal H}$, $\wt{\cal C}$, and $\wt{\cal N}$ 
satisfying these relations. First of all,
the last condition in (\ref{constraint-eqnss}) 
shows that $\wt {\cal N}$ and $\wt {\cal C}$ are not independent variables.
A useful way to express this dependence  is to introduce a new $O(d,d)$ vector
${\cal C}$ via the equation
\be \label{ecwc}
\wt {\cal C} \ = \ {\cal C} (1 + \kappa^{-1}\, \wt {\cal N}) \,. 
\ee
Indeed, the equation leads to
   \be\label{DetN}
   (1+\wt{{\cal N}}\kappa^{-1}) {\cal C}^{\,T}\eta\,  {\cal C}\, (1+\kappa^{-1} \wt{{\cal N}}) \ = \ 
   \kappa-\wt{\cal N}\kappa^{-1}\wt{\cal N}
   \ = \ 
   \kappa(1-\kappa^{-1}\,\wt{{\cal N}})(1+\kappa^{-1}\,\wt{{\cal N}})\;, 
  \ee 
giving us  
 \be\label{tildeNREL}
 \wt {\cal N} \ = \ (\kappa- {\cal C}^{\,T} \eta  {\cal C}) (\kappa +  {\cal C}^{\,T} \eta {\cal C})^{-1}\kappa
 \ = \ \kappa \,  (\kappa +  {\cal C}^{\,T} \eta {\cal C})^{-1}  (\kappa- {\cal C}^{\,T} \eta  {\cal C})  \, \, .
 \ee
Eqs.~\eqref{tildeNREL} and \eqref{ecwc} express both $\wt{\cal N}$ and $\wt{\cal C}$ in terms of ${\cal C}$.
For later use we note that  
 \be\label{usefulREL}
  \kappa^{-1}+\kappa^{-1}
  \wt{{\cal N}}\kappa^{-1} \ = \ \kappa^{-1}
  \Big[(\kappa+ {\cal C}^{\,T}\eta\, {\cal C})+(\kappa- {\cal C}^{\,T}\eta\, {\cal C})\Big]
  (\kappa+ {\cal C}^{\,T}\,\eta {\cal C})^{-1}\ = \ 2(\kappa+ {\cal C}^{\,T}\eta 
  {\cal C})^{-1}\;. 
 \ee

We now introduce an $O(d,d)$ valued generalized metric and an $O(d,d)$ vector 
to parametrize the above fields.
We claim that the first two conditions in \eqref{constraint-eqnss} can be solved by
taking
\be \label{eclaim}
\wt {\cal H} \ = \   {\cal H} + {\cal C} (\kappa^{-1}+ \kappa^{-1}\wt {\cal N}\kappa^{-1}) {\cal C}^{\,T} \,,
\ee
where ${\cal H}$ is a
symmetric matrix satisfying:\footnote{
The second condition is a projector
condition because $P \equiv \tfrac{1}{2} (1 + {\cal H}\eta)$ satisfies
$P^2 = P$.  The complementary orthogonal projector is $\bar P \equiv \tfrac{1}{2} (1 - {\cal H}\eta)$.  We now explain that the alternative
condition $(1 - {\cal H}\eta){\cal C} =0$ would not be viable.  Consider the
second constraint in \refb{constraint-eqnss}, 
$\wt {\cal H}\, \eta \, \wt {\cal C} + \wt {\cal C} \,\kappa^{-1} \wt {\cal N} \ = 0$, perturbatively around zero ${\cal C}$. 
To leading order we have $\wt{\cal N} = \kappa$ and we require 
$\wt{\cal C} = {\cal C}$ 
and $\wt{\cal H} = {\cal H}$.  The constraint becomes 
$({\cal  H} \eta + 1) {\cal C} =0$.   The choice of this projector was
fixed by our convention for the duality group and its associated
metric.  We picked $O(d , d+ K)$ where $d+K$ is the number of positive
eigenvalues of $\wh\eta$, including the positive eigenvalues of $\kappa$.
Had we chosen $O(d+K, d)$,  we would have to change $\kappa \to -\kappa$ in $\wh\eta$ and the other projector would have been selected.}
\be \label{epp4s}
{\cal H}\, \eta\,  {\cal H} \ = \ \eta\,  , \quad (1+{\cal H}\,\eta) 
\, {\cal C} \ = \ 0\, .
\ee
This can be easily verifed by substituting \refb{eclaim} into the first two
equations of  \eqref{constraint-eqnss} and using
\refb{epp4s}, \eqref{ecwc}, and \eqref{tildeNREL}. 
Furthermore since ${\cal H}$ is determined
uniquely from $\wt{\cal H}$, $\wt{\cal C}$, and $\wt{\cal N}$ 
using eqs.\eqref{eclaim} and \refb{ecwc}, 
\refb{eclaim} is the most general form of $\wt{\cal H}$
satisfying \refb{constraint-eqnss}.

These results can now by summarized in the statement 
that $\wt{\cal H}$, $\wt{\cal C}$, and $\wt{\cal N}$ satisfying 
\refb{constraint-eqnss} can be parametrized
by ${\cal H}$ and ${\cal C}$ satisfying \eqref{epp4s} via the relations 
\be \label{ehexp9}
\begin{split}
\wt {\cal H} \ = \  & \ {\cal H} + {\cal C} (\kappa^{-1}+ \kappa^{-1}\wt {\cal N}\kappa^{-1}) {\cal C}^{\,T} \,, \\
\wt {\cal C}   \ = \ & \ {\cal C}\,  (1 + \kappa^{-1} \wt {\cal N})\;,  \ \\
 1+\wt{{\cal N}}\kappa^{-1} \ = \ & \ 2\kappa\, (\kappa+ {\cal C}^{\,T} \eta {\cal C})^{-1}\,.
\end{split}
\ee
Alternatively, using (\ref{usefulREL}), this can be written as 
 \be
  \begin{split}\label{ehexp9II}
   \wt {\cal H} \ = \  & \ {\cal H} + 2\,{\cal C} (\kappa+ {\cal C}^{T}\eta {\cal C})^{-1} {\cal C}^{\,T} \,, \\
\wt {\cal C}   \ = \ & \ 2\,{\cal C}\,  (\kappa + {\cal C}^{T}\eta {\cal C})^{-1}\kappa \;, \\
 \wt{{\cal N}} \ = \ & \ 
 -\kappa+2\,\kappa\left(\kappa+{\cal C}^T\eta {\cal C}\right)^{-1}\kappa
 \,.
\end{split}
\ee

Our next goal is to write the proper $O(d,d)$ covariant  
objects ${\cal H}$ and ${\cal C}$ in terms of the 
physical fields.\footnote{Ref.~\cite{Jeon:2011kp} has 
examined the use of unconstrained $O(d,d)$ vectors for
a DFT of Yang-Mills fields. }  
Thus, consider the expressions for 
$\wt {\cal H}$, $\wt {\cal C}$, and $\wt {\cal N}$ in terms of the dimensionally reduced physical variables given in 
(\ref{newH}). Using matrix notation for $G$, $B$, and $(a)^{\alpha}{}_{i}\equiv a_{i}{}^{\alpha}$
we read off 
\ben
\label{hcn-ms}
\wt {\cal H} &=& \begin{pmatrix}
G^{-1} & -G^{-1}\left(B + {1\over 2} a^T\kappa a\right)\phantom{\Bigl(} 
\\[0.8ex] 
\  \left(\ B - {1\over 2} a^T \kappa a\right)G^{-1} & \ \   G + \left(-B + {1\over 2} a^T \kappa a\right) G^{-1}
\left(B + {1\over 2} a^T\kappa  a\right) + a^T\kappa a 
\end{pmatrix}  \,, \nonumber \\[0.2ex]
\wt {\cal C} &=& \begin{pmatrix} - G^{-1} a^T\kappa \phantom{\Bigl(}\cr \left(-B + {1\over 2} a^T\kappa a\right) 
G^{-1} a^T\kappa + a^T\kappa
\end{pmatrix}\ = \ \begin{pmatrix} - G^{-1} a^T\kappa \phantom{\Bigl(}\cr \left(-B+\bar G\right) G^{-1} a^T\kappa 
\end{pmatrix} \,,   \nonumber \\[0.5ex]
\wt {\cal N} &=& \kappa + \kappa\, a G^{-1} a^T\,\kappa\, .
\een
Here we defined 
 \be
 \label{coldnew} 
   \bar G \ \equiv \ G+\tfrac{1}{2}\, a^T\kappa\,  a\; ,
 \ee
for later convenience.  
Using this we can now express the new field variables ${\cal H}$ and ${\cal C}$ 
in terms of the
 physical fields.  We begin with  ${\cal C}$: 
  \be
    {\cal C} \ = \ \wt {{\cal C}}(1+ \kappa^{-1}\wt{{\cal N}})^{-1} \ = \ 
   \begin{pmatrix} - G^{-1} a^T\kappa \phantom{\Bigl(}\cr (-B+\bar G) G^{-1} a^T\kappa
\end{pmatrix} \tfrac{1}{2}\big(1+\tfrac{1}{2} \, a\,G^{-1}a^T\kappa\big)^{-1}\;.
  \ee
Working out the geometric series one finds that this can be written in terms of the redefined metric $\bar G$:
   \be
   \label{cnew} 
      {\cal C} \  = \ 
   \tfrac{1}{2}
   \begin{pmatrix} 
   - \bar {G}^{\,-1} a^T\kappa \phantom{\Bigl(}\cr (-B+\bar G) \bar{G}^{\,-1}a^T\kappa
   \end{pmatrix} 
= \  \tfrac{1}{2}
   \begin{pmatrix} - \bar{G}^{\, -1} a^T\kappa  \phantom{\Bigl(}\cr -B 
   \bar{G}^{\, -1} a^T\kappa + a^T\kappa 
   \end{pmatrix}\;. 
 \ee  
Next we turn to ${\cal H}$. 
From (\ref{ehexp9}) and the last of (\ref{hcn-ms}) we have    
\be 
 {\cal H} \ = \    \wt {\cal H} - {\cal C} (\kappa^{-1} + \kappa^{-1}\wt {\cal N}\kappa^{-1}) \,
 {\cal C}^{\,T} \ = \ \wt {\cal H} - \,2\, {\cal C} \bigl(\kappa^{-1}+ 
 \tfrac{1}{2} \, a \, G^{-1} a^T\bigr) {\cal C}^T \,.\ee     
Using our expression for ${\cal C}$ in terms of the physical fields and
that for $\wt {\cal H}$,  
an explicit computation gives a very simple result
for ${\cal H}$:
\be
\label{h-physical}
{\cal H}\ = \ \begin{pmatrix}
\bar G^{\,-1} & -\bar G^{\,-1}  B\phantom{\Bigl(}\cr
B\,\bar G^{\,-1}   & \bar G\ - \ B\,\bar G^{\,-1}\,B
\end{pmatrix}  \,. 
 \ee
This $O(d,d)$ valued generalized metric ${\cal H}$ takes
the usual form with the metric 
$G$ replaced by $\bar G$.  
It is a good exercise to verify that the constraint on ${\cal C}$ holds:
$(1+ {\cal H}\eta) {\cal C} =0$. 

Since ${\cal H}$ transforms in the familiar way under $O(d,d)$ dualities, the
fields $\bar G$ and $B$ transform in the familiar way. Since the
transformation of $\bar G$ is known and ${\cal C}$ transforms
as an $O(d,d)$ vector, this determines the duality transformation
of the 
scalar fields $a$.  Given this, one can find the
duality transformations of $G$.

Assembling the $O(d,d)$ multiplets for the gauge fields  requires no work.
Recalling equations \refb{gfdef78} and \refb{fsdef67} we write now
\be
 \label{gfdef78pp}
 \begin{split}
 \wh  A_{\mu}{}^{\hat{M}} \ \equiv  \ & \ \big(\,A_{\mu }{}^M\,,\; \;A_{\mu}{}^{\alpha}\,\big) \;, \\
 \wh  {\cal F}_{\mu\nu}{}^{\hat{M}} \ \equiv \ & \ \big(\,{\cal F}_{\mu\nu}{}^M \,,\; {\cal F}_{\mu\nu}{}^{\alpha}\,\big)\;. 
 \end{split}
 \ee 
The field strengths are computed in terms of the gauge fields as 
\be
\label{fintez9}
\begin{split}
 {\cal F}_{\mu\nu}{}^M\ & \equiv \  \partial_\mu {A}_{\nu}{}^M
 - \partial_\nu {A}_{\mu}{}^M\,, \\ 
{\cal F}_{\mu\nu}{}^\alpha 
  \ &\equiv  \ 2\, \partial_{[\mu}A_{\nu]}{}^\alpha 
  +f^\alpha{}_{\beta\gamma}A_{\mu}{}^{\beta}
  A_{\nu}{}^{\gamma}\, .
\end{split}
\ee
  
\subsection{Covariant action}

 We now treat ${\cal H}$ and ${\cal C}$ as independent variables and formulate the action in terms of these fields.  
 Since the action has an explicit expression in terms of $\wt {\cal H}$, 
 $\wt {\cal C}$, and
 $\wt {\cal N}$, which in turn have known expressions in terms of ${\cal H}$ and ${\cal C}$,
 the action is guaranteed to have an explicit expression in terms of 
 ${\cal H}$ and ${\cal C}$. 

We start with the scalar kinetic terms from (\ref{hetACTION}),  
\be
L_{\rm kin} \ = \   \tfrac{1}{8}  
D_\mu \wh{\cal H}^{\hat M\hat N} D^\mu  
\wh {\cal H}_{\hat M\hat N}  
\ = \ \tfrac{1}{8}  \, \hbox{tr}  \Bigl[\, (D_\mu \wh{\cal H}) \,\wh \eta^{\,-1}   (D^\mu \wh {\cal H} ) \wh\eta^{\,-1} \,  \Bigr]  \;, 
\ee
where $\wh \eta$ is the $O(d,d+K)$ metric (\ref{neweta}). 
Using (\ref{epp1}) and expanding the blocks,  
\be\label{EXPandingBLock}
L_{\rm kin} 
 \ = \  \tfrac{1}{8} \, \hbox{Tr} 
 \bigl(  D_\mu \wt {\cal H}  \, \eta \, D^\mu \wt {\cal H}\,  \eta  
 + 2 D_\mu \wt {\cal C}^{\,T} \eta \, 
D^\mu \wt {\cal C}  \kappa^{-1}  
+  D_\mu \wt {\cal N}  \kappa^{-1} D^\mu \wt 
{\cal N}  \kappa^{-1} \bigr) \;. 
\ee
Next we insert (\ref{ehexp9}) and use 
(\ref{tildeNREL}) to simplify these terms. 
The strategy is to rewrite all terms so that only derivatives of ${\cal C}$ enter. 
To this end one uses  
the second constraint in (\ref{epp4s}) to find  
 \be
  {\cal C}^T\eta D_{\mu}{\cal H} \ = \ -D_{\mu}{\cal C}^T(1+\eta{\cal H})\;, 
 \ee
which allows us to eliminate derivatives of 
${\cal H}$ (note that $D_\mu {\cal H} = \partial_\mu {\cal H}$), 
 and 
the third equation in (\ref{ehexp9}) to find
 \be
  D_{\mu}(1+\wt{\cal N}\kappa^{-1}) \ = \ -2\kappa(\kappa+{\cal C}^T\eta{\cal C})^{-1}
  (D_{\mu}{\cal C}^T\eta {\cal C}+{\cal C}^{T}\eta D_{\mu}{\cal C})(\kappa+{\cal C}^T\eta{\cal C})^{-1}\;, 
  \ee
which allows us to eliminate derivatives of $\wt{\cal N}$.   
A direct but  tedious computation then shows that  (\ref{EXPandingBLock}) can be written in the form 
 \be\label{kinTERM}  
 \begin{split}
  { L}_{\rm kin} \ = \ \ {\rm Tr}\ \Big[&\  
  \tfrac{1}{8}\,\eta\partial^{\mu}{\cal H}\eta\,\partial_{\mu} {\cal H}
  -D^{\mu}{\cal C}\big(\kappa+{\cal C}^T\eta {\cal C})^{-1}D_{\mu} {\cal C}^T\eta{\cal H}\eta\\
  & -D^{\mu}{\cal C}^T\eta {\cal C}\big(\kappa+{\cal C}^T\eta {\cal C}\big)^{-1}{\cal C}^T\eta 
  D_{\mu}{\cal C}\big(\kappa+{\cal C}^T\eta {\cal C}\big)^{-1}\,\Big]\;. 
 \end{split}
 \ee 
This action can be written in various equivalent forms, some of which that may be more illuminating  are given   
in the following. Using that with the constraints (\ref{epp4s}) we 
have ${\cal H}^{-1} = \eta {\cal H} \eta$  and hence 
 \be
  {\cal C}^T \eta {\cal C} \ = \ -{\cal C}^T\eta{\cal H}\eta{\cal C} \ = \ - {\cal C}^T {\cal H}^{-1} {\cal C}\;, 
 \ee 
we can write the action, upon cycling in the trace, as
 \be\label{kinTERMv}  
 \begin{split}
  {L}_{\rm kin} \ = \ \ {\rm Tr}\ \Big[&\  
  \tfrac{1}{8}\, \partial_{\mu}{\cal H}\, \partial^{\mu} {\cal H}^{-1} 
  -\,{\cal H}^{-1} D_{\mu}{\cal C}\big(\kappa-{\cal C}^T{\cal H}^{-1} {\cal C})^{-1} D^{\mu} {\cal C}^T \\
  & -\, {\cal H}^{-1} {\cal C}
  \big(\kappa-{\cal C}^T {\cal H}^{-1} {\cal C}\big)^{-1}{\cal C}^T{\cal H}^{-1}\,  D_{\mu}{\cal C}
  \big(\kappa-{\cal C}^T {\cal H}^{-1}  {\cal C}\big)^{-1}\,D^{\mu}{\cal C}^T\ \Big]\;. 
 \end{split}
 \ee 
Next we can group the last two terms as follows   
\be\label{kinTERMvv}  
  {L}_{\rm kin} \ = \ {\rm Tr}\ \Big[ 
  \tfrac{1}{8}\, \partial_{\mu}{\cal H}\, \partial^{\mu} {\cal H}^{-1} 
   -\, \Bigl( {\cal H}^{-1} +{\cal H}^{-1} {\cal C}  (\kappa-{\cal C}^T {\cal H}^{-1}  {\cal C})^{-1} {\cal C}^T 
   {\cal H}^{-1}  \Bigr) \,  D_{\mu}{\cal C}\big(\kappa-{\cal C}^T {\cal H}^{-1}  {\cal C}\big)^{-1}\,
   D^{\mu}{\cal C}^T\Big]\;. 
\ee
The prefactor in the second term can be simplified, as one may verify by writing out the 
geometric series, to obtain   
\be\label{kinTERMvvv}  
 {L}_{\rm kin} \ = \ {\rm Tr}\ \Big[ 
  \tfrac{1}{8}\, \partial_{\mu}{\cal H}\, \partial^{\mu} {{\cal H}}^{-1}  -\,  ({\cal H}-{\cal C}\kappa^{-1} 
  {\cal C}^{\,T})^{-1}      D_{\mu}
  {\cal C}\; (\kappa-{\cal C}^{\,T} \,{\cal H}^{-1} {\cal C})^{-1} D^{\mu}
  {\cal C}^{\,T} \Big]\;. 
 \ee 
This form of the kinetic terms 
makes it clear that the variables $\CC_{M\alpha}$ have 
restricted domain since the eigenvalues of $({\cal H}-{\cal C}\kappa^{-1} 
  {\cal C}^{\,T})$ and $(\kappa-{\cal C}^{\,T} \,{\cal H}^{-1} {\cal C})$ should never vanish, and
  hence, by the positivity of these eigenvalues at $\CC=0$, must always remain positive.
  By writing $\HH = AA^T$ and $\kappa = B^TB$ for some non-singular matrices $A$, $B$ we can translate both these conditions into positivity of the eigenvalues of
  \be
  (I_{2d} - \bar \CC \, \bar \CC^T)\,, \qquad \bar \CC \ \equiv \ A^{-1} \CC \, B^{-1} \, .
  \ee
This means that the eigenvalues of $\bar\CC \, \bar\CC^T$ should 
be less than one.
A particular consequence of this is that $\hbox{Tr}(\bar \CC \, \bar\CC^T)< 2d$. Since 
the left hand side is the sum of squares of all the 
components of $\bar\CC$, 
this shows that each
component of $\bar\CC$ 
has a strict upper bound. Thus, for fixed $A$ and $B$,
i.e.\ fixed $\HH$ and $\kappa$, each component of $\CC$ will also have a strict
upper bound. 
This does not impose, however,  
 any condition on the physical fields $a_m{}^\alpha$. 
To see this we note that using the constraint $(1+ {\cal H} \eta) {\cal C} =0$, the third equation in \refb{ehexp9} 
and the third equation in
\refb{hcn-ms} we find
\be
\begin{split}
\label{scomp}
(\kappa-{\cal C}^{\,T} \,{\cal H}^{-1} {\cal C})^{-1} \ = \    
(\kappa+{\cal C}^{\,T} \eta \,{\cal C})^{-1} \ = \ \tfrac{1}{2} \kappa^{-1} (1 + \wt{\cal N} \kappa^{-1} ) 
\ = \   \kappa^{-1} + \tfrac{1}{2} \, aG^{-1} a^T  \,.
\end{split}
\ee
This shows that the inverse matrix on the left-hand side always exists
for finite $a_m{}^\alpha$ and that the bounds on ${\cal C}$ do not
impose extraneous conditions.

Let us now turn to the potential for the scalar fields, 
which can be obtained from 
 \be
  V \ = \ f^{\alpha\beta\gamma} f^{\alpha'\beta'\gamma'}\Big[\,
  \tfrac{1}{12} \wt{\cal N}_{\alpha\alpha'} \wt{\cal N}_{\beta\beta'} \wt{\cal N}_{\gamma\gamma'}
  -\tfrac{1}{4}\kappa_{\alpha\alpha'}\kappa_{\beta\beta'}\wt {\cal N}_{\gamma\gamma'}
  +\tfrac{1}{6}\kappa_{\alpha\alpha'}\kappa_{\beta\beta'}\kappa_{\gamma\gamma'}\, \Big]\; .
 \ee
Upon replacing $\wt{\cal N}$ with the second equation in (\ref{tildeNREL}) 
one may verify that the potential can be brought into the form  
\ben\label{CPOTpre}  
V &=& f_{\alpha\beta\gamma} f^{\alpha'\beta'\gamma'} \left[(\kappa+
{\cal C}^{\,T} \eta {\cal C})^{-1} 
\right]^{\alpha\alpha''} 
 \left[(\kappa+
{\cal C}^{\,T} \eta {\cal C})^{-1} 
\right]^{\beta\beta''}  \left[(\kappa+
{\cal C}^{\,T} \eta {\cal C})^{-1} 
\right]^{\gamma\gamma''}
\nonumber \\
&&
\left[ \tfrac{1}{12}(\kappa-
{\cal C}^{\,T} \eta {\cal C})_{\alpha''\alpha'} (\kappa-
{\cal C}^{\,T} \eta {\cal C})_{\beta''\beta'} (\kappa-
{\cal C}^{\,T} \eta {\cal C})_{\gamma''\gamma'} \right.\nonumber \\ && \left.
-\tfrac{1}{4}(\kappa+
{\cal C}^{\,T} \eta {\cal C})_{\alpha''\alpha'} (\kappa+
{\cal C}^{\,T} \eta {\cal C})_{\beta''\beta'} (\kappa-
{\cal C}^{\,T} \eta {\cal C})_{\gamma''\gamma'} \right. \nonumber \\ && \left.
+\tfrac{1}{6}(\kappa+
{\cal C}^{\,T} \eta {\cal C})_{\alpha''\alpha'} (\kappa+
{\cal C}^{\,T} \eta {\cal C})_{\beta''\beta'} (\kappa +
{\cal C}^{\,T} \eta {\cal C})_{\gamma''\gamma'}\right]\, .
\een
Expanding and simplifying the terms inside the last square bracket and relabelling the indices
in some terms we get
\ben\label{CPOT}  
V &=& f_{\alpha\beta\gamma} f^{\alpha'\beta'\gamma'} \left[(\kappa+
{\cal C}^{\,T} \eta {\cal C})^{-1} 
{\cal C}^{\,T}\eta \,{\cal C}\right]^{\alpha}{}_{\alpha'} \left[ (\kappa+
{\cal C}^{\,T} \,\eta\,{\cal  C})^{-1} 
{\cal C}^{\,T}\eta\, {\cal C}\right]^{\beta}{}_{\beta'}  
\nonumber \\ &&
\qquad \qquad\quad  \cdot \left[ (\kappa+{\cal C}^{\,T} \eta {\cal C})^{-1} 
\left( \kappa + \tfrac{1}{3} {\cal C}^{\,T}\eta {\cal C}\right)\right]^{\gamma}{}_{\gamma'}\; .
\een
This form makes it manifest that the potential has no constant terms, i.e., there 
is no cosmological constant, and no terms quadratic in ${\cal C}$, i.e., 
there are no mass terms for vacua in which the scalars have zero expectation value. 

Finally, we rewrite the Yang-Mills term using the $O(d,d)$ covariant field variables.  
Starting  from 
\be\label{LYM}  
L_{\rm YM} \ = \ -\tfrac{1}{4}\,  \wh {\cal H}_{\hat M \hat N}\,   
\wh {\cal F}^{\mu\nu \hat M}\, 
\wh {\cal F}_{\mu\nu}{}^{\hat N} \;,
\ee
 we insert the block components of $\wh{\cal H}$ according to (\ref{ehexp9II}) 
and the components of the field strengths $\wh {\cal F}$ 
according to \refb{gfdef78pp}. 
The resulting Yang-Mills term is simplified by introducing the following 
combination,
 \be
  \overline{\cal F}_{\mu\nu}{}^{M} \ \equiv \ {\cal F}_{\mu\nu}{}^{M}-\eta^{MN}C_{N\alpha}{\cal F}_{\mu\nu}{}^{\alpha}\;. 
 \ee 
A straightforward computation then shows that (\ref{LYM}) can be written as 
 \be
 \begin{split}
  L_{\rm YM} \ = \ \,&-\tfrac{1}{4}\,
  \big({\cal H}+2\,{\cal C}(\kappa+{\cal C}^T\eta{\cal C})^{-1}{\cal C}^{T}\big)_{MN}\,
  \overline{\cal F}^{\mu\nu \,M}\,\overline{\cal F}_{\mu\nu}{}^{N}
  -\tfrac{1}{2}\,{\cal C}_{M\alpha}\,\overline{\cal F}^{\mu\nu\,M}\, {\cal F}_{\mu\nu}{}^{\alpha} \\[0.4ex]
  &-\tfrac{1}{4}\left(\kappa+{\cal C}^T\eta{\cal C}\right)_{\alpha\beta}\,{\cal F}^{\mu\nu\,\alpha}\,{\cal F}_{\mu\nu}{}^{\,\beta}\;. 
 \end{split}
 \ee 
 By using $ \eta \, {\cal C}=-{\cal H}^{-1}{\cal C} $ and  
 expanding and resumming the geometric series we can also rewrite this as
  \be
 \begin{split}
  L_{\rm YM} \ = \ \,&-\tfrac{1}{4}\,
  \big({\cal H} \, ({\cal H} - {\cal C} \kappa^{-1}
   {\cal C}^T)^{-1}   
  ({\cal H} + {\cal C} \kappa^{-1} 
  {\cal C}^T) \big)_{MN}\,    
    \overline{\cal F}^{\mu\nu \,M}\,\overline{\cal F}_{\mu\nu}{}^{N}
  -\tfrac{1}{2}\,{\cal C}_{M\alpha}\,\overline{\cal F}^{\mu\nu\,M}\, {\cal F}_{\mu\nu}{}^{\alpha} \\[0.4ex]
  &-\tfrac{1}{4}\left(\kappa+{\cal C}^T\eta\, {\cal C}\right)_{\alpha\beta}\,{\cal F}^{\mu\nu\,\alpha}\,{\cal F}_{\mu\nu}{}^{\,\beta}\;. 
 \end{split}
 \ee

We are now ready to assemble the pieces and give the final form of the dimensionally 
reduced action, thus summarizing our result.  
 The action is  written in terms of the field content 
 \be
  \{\; g_{\mu\nu}\,,\; b_{\mu\nu}\,,\;\phi \,,\;  A_{\mu}{}^{\alpha}\,,\; A_{\mu}{}^{M}\,,\; 
  {\cal H}_{MN}\,,\;  {\cal C}_{M}{}^{\alpha}\;\}\;. 
 \ee
Here the first four fields are $O(d,d)$ singlets, and the final three fields transform
in $O(d,d)$ tensor representations.  
We use matrix notation: ${\cal H}$ for ${\cal H}_{MN}$
 and ${\cal C}$ for ${\cal C}_{M\alpha}$.  The matrix ${\cal H}$ satisfies the
 familiar constraint of the generalized metric ${\cal H} \,\eta\, {\cal H} = \eta$,
 while ${\cal C}$ is constrained 
 by $(1+ {\cal H}\eta)\,  {\cal C} =0$. 
In terms of these variables
 and using matrix notation  the action reads  
   \be\label{hetACTIONFINAL}  
  \begin{split}
   S \ = \ \int d^nx\,
   \sqrt{-g}  
   \,&e^{-2\phi}\Big(R(g) +4\,\partial^{\mu}\phi\,\partial_{\mu}\phi
   -\tfrac{1}{12}H^{\mu\nu\rho}H_{\mu\nu\rho}  
   +\tfrac{1}{8}\, {\rm Tr} \big( \,   
   \partial^{\mu}{\cal H}\, \partial_{\mu} {{\cal H}}^{-1}\big)  \\
   &-{\rm Tr} \Big[ ({\cal H}-{\cal C}\kappa^{-1} 
  {\cal C}^{\,T})^{-1}      D^{\mu}
  {\cal C}\; 
  {\cal K}_{\cal C}^{-1} 
  D_{\mu}
  {\cal C}^{\,T} \Big]   -V({\cal C})    \\[0.5ex]
   &-\tfrac{1}{4}
  \big( {\cal H} \, ({\cal H} - {\cal C} \kappa^{-1} {\cal C}^T)^{-1}
  ({\cal H} + {\cal C} \kappa^{-1} {\cal C}^T)\big)_{MN}\,
    \overline{\cal F}^{\mu\nu \,M}\,\overline{\cal F}_{\mu\nu}{}^{N}\\[0.5ex]
  &-\tfrac{1}{2}\,{\cal C}_{M\alpha}\,\overline{\cal F}^{\mu\nu\,M}\, {\cal F}_{\mu\nu}{}^{\alpha}
   -\tfrac{1}{4}
  ( {\cal K}_{\cal C})_{\alpha\beta}\,{\cal F}^{\mu\nu\,\alpha}\,
   {\cal F}_{\mu\nu}{}^{\,\beta}\Big)\;, 
  \end{split}
 \ee  
where we defined the ${\cal C}$ dependent extension ${\cal K}_{\cal C}$ of the Cartan-Killing metric: 
\be
{\cal K}_{\cal C} \ \equiv \ \kappa-{\cal C}^{\,T} \,{\cal H}^{-1} {\cal C}\ = \ 
\kappa+{\cal C}^T\eta{\cal C}\,.
\ee
The potential $V({\cal C})$ 
is given by (\ref{CPOT}), 
\ben\label{CPOTb}  
V \hskip-5pt &=& \hskip-5pt f_{\alpha\beta\gamma} f^{\alpha'\beta'\gamma'} \hskip-2pt
\left[({\cal K}_{\cal C})^{-1} 
{\cal C}^{\,T}\eta \,{\cal C}\right]^{\alpha}{}_{\alpha'} 
\left[ ({\cal K}_{\cal C})^{-1} 
{\cal C}^{\,T}\eta\, {\cal C}\right]^{\beta}{}_{\beta'}  
 \left[ ({\cal K}_{\cal C})^{-1} 
\left( \kappa + \tfrac{1}{3} {\cal C}^{\,T}\eta {\cal C}\right)\right]^{\gamma}{}_{\gamma'}\; , 
\een
 the 3-form curvature takes the form  
 \be\label{Recall3form}
  H_{\mu\nu\rho} \ = \ 3\Big(\partial_{[\mu}\, b_{\nu\rho]}-A_{[\mu}{}^{M}\partial^{}_{\nu}A_{\rho] M}  
   -{A}_{[{\mu}}{}^{\alpha}\partial^{}_{{\nu}}{A}_{{\rho}]\,\alpha} 
   -\tfrac{1}{3}f_{\alpha\beta\gamma}{A}_{{\mu}}{}^{\alpha}{A}_{{\nu}}{}^{\beta}
    {A}_{{\rho}}{}^{\gamma}\Big)\,,
 \ee 
 and the field strengths and covariant derivatives are  
  \be\label{nonABELian}
 \begin{split}
  \overline{\cal F}_{\mu\nu}{}^{M} \ & \equiv \ {\cal F}_{\mu\nu}{}^{M}-\eta^{MN}C_{N\alpha}{\cal F}_{\mu\nu}{}^{\alpha}\,, \\
 {\cal F}_{\mu\nu}{}^{M}\ & \equiv \  \partial_\mu {A}_{\nu}{}^{M} 
 - \partial_\nu {A}_{\mu}{}^{M}\,, \\
   {\cal F}_{\mu\nu}{}^\alpha 
  \ &\equiv  \ 2\partial_{[\mu}A_{\nu]}{}^\alpha 
  +f^{\alpha}{}_{\beta\gamma}A_{\mu}{}^{\beta}
  A_{\nu}{}^{\gamma}\, , \\ 
  D_\mu {\cal C}_{M\alpha} \ & \equiv \ \partial_\mu{\cal C}_{M\alpha}
  - A_\mu{}^\gamma \, f_{\gamma\alpha}{}^\beta {\cal C}_{M\beta}\,.  
 \end{split}
 \ee  

This form of the action is manifestly $O(d,d)$ invariant as it is written in terms of 
$O(d,d)$ covariant objects, with all indices properly contracted. 
Since both ${\cal H}$ and ${\cal C}$ are constrained, an unconstrained
parameterization of these objects is useful. 
 If we parameterize the matrix ${\cal H}$ 
 using a symmetric matrix of scalars $\bar G$ and an antisymmetric
matrix of scalars $B$:
\be
\label{h-physical--}
{\cal H}\ = \ \begin{pmatrix}
\bar G^{\,-1} & -\bar G^{\,-1}  B\phantom{\Bigl(}\cr
B\,\bar G^{\,-1}   & \bar G\ - \ B\,\bar G^{\,-1}\, B
\end{pmatrix}  \,, 
 \ee
then we can give also an explicit parameterization of ${\cal C}$
in terms of a field $A\equiv A_{m\alpha}$:
  \be
      {\cal C} \  = \  \tfrac{1}{2}
   \begin{pmatrix} - \bar {G}^{\, -1} A  \phantom{\Bigl(}\cr -B\,  \bar {G}^{\, -1} A + A
   \end{pmatrix}\;. 
 \ee  
We can then view $\bar G, B,$ and $A$ as independent fields.
The connection to the 
original supergravity variables  
yields a slightly different and more complex  
parameterization in which   
$A$ above is set equal  to 
$a^T\kappa$ and $ \bar G$ is set equal
to $  G+\tfrac{1}{2}\, a^T\kappa a$, 
 as one can recall from \refb{coldnew} and \refb{cnew}. 

The results of this section apply also if the original higher-dimensional 
gauge group is of the form $G'\times U(1)^p$, with $G'$ compact semisimple and of dimension $K'$. As we discussed at the end of the previous section, the 
 reduced theory is formally $O(d, d+p+K')$ invariant but 
the true symmetry is only $O(d,d+p)$.  This time we want to write
the theory in terms of 
$O(d, d+p)$ multiplets. 
The analysis in this case is a straightfoward 
generalization of the analysis of this section.  Following the
discussion of this situation around \refb{newhmatrix}, the matrix $\wh{\cal H}$ in \refb{epp1explicit}  takes now a similar form 
\be \label{epp1explicit99}
 \widehat {\cal H}_{\hat M \hat N} \ = \ \begin{pmatrix} \wt {\cal H}_{MN} & \wt {\cal C}_{M\beta}  \\[0.4ex]
  (\wt {\cal C}^{\,T})_{\alpha N}  & \wt {\cal N}_{\alpha\beta} \end{pmatrix} \, ,
 \ee
but the matrix
 dimensions are now as follows 
 \be
 \label{sizes9} 
 \begin{split}
 \wh \eta \ ,\  \wh{\cal H} \ : \  & \  (2d+p+K') \times (2d+p+ K') \,,\\
 \eta \ ,\ \wt{\cal H} \ : \  & \  (2d+ p) \times (2d+ p) \, , \\
 \wt{\cal C} \ : \  & \  (2d+p) \times K' \,,  \\
\kappa'\ , \  \wt{\cal N} \ : \  & \  K' \times K'  \,.  
 \end{split}
 \ee 
$\eta$ will now
correspond to the matrix 
$$\eta \ =  \ \begin{pmatrix} 0 & I_d & 0\cr I_d & 0 & 0\cr 0 & 0 & I_{p}
\end{pmatrix}\,. $$ 
The analysis
proceeds as before.  We parameterize $\wt{\cal H}, \wt{\cal C}$, and 
$\wt{\cal N}$, with matrices ${\cal H}$ and ${\cal C}$ of  sizes 
\be
 \label{sizes9aa}  
 \begin{split}
 {\cal H} \ : \  & \  (2d+ p) \times (2d+ p)\,, \\
{\cal C} \ : \  & \  (2d+p) \times K' \,, 
 \end{split}
 \ee 
that
satisfy the same constraints as before
(${\cal H}\, \eta\,  {\cal H} = \eta\,, \  (1+{\cal H}\,\eta) 
\, {\cal C}  =  0$), and thus have the same solutions. 
As a result, the 
action takes the  form \eqref{hetACTIONFINAL} with ${\cal H}\,\eta^{-1}$ an $O(d,d+p)$ matrix and ${\cal C}$ 
an $O(d,d+p)$ vector valued in the Lie algebra of $G'$. Finally, $\kappa$ is set equal to
the Cartan-Killing metric $\kappa'$ of the Lie algebra of $G'$.

 \section{Conclusions}
 
In this paper we have revisited the effective action of heterotic string theory on 
a torus and its duality symmetries. The seminal work of Maharana-Schwarz 
exhibited a global $O(d,d+16;\mathbb{R})$ symmetry in the reduction of 
heterotic supergravity truncated to the Cartan subalgebra on a $d$-torus.
This is the proper effective theory for the situation in which the background fields 
corresponding to the metric $g$, $b$-field and the gauge fields of the Cartan subalgebra 
all have non-trivial 
values. In fact, in this case the non-Cartan parts of the 
gauge groups $SO(32)$ or $E_8\times E_8$ are `Higgsed' and hence massive. 
Therefore they have to be ignored in the massless effective action, giving the theory 
with global $O(d,d+16;\mathbb{R})$ symmetry constructed by Maharana-Schwarz.
However, we may also consider the situation for which only $g$ and $b$ have non-trivial 
background values. In this case the full gauge fields corresponding to $SO(32)$ or $E_8\times E_8$ 
remain massless, and so the question arises what is the global duality symmetry 
upon including all these non-abelian gauge fields. We investigated this question, 
showed that the duality symmetry is $O(d,d;\mathbb{R})$ in general and 
exhibited this symmetry in the novel effective action (\ref{hetACTIONFINAL}) in 
manifest form. 
Interestingly, such a formulation requires non-polynomial couplings in the $O(d,d)$
covariant fields as is manifest, for instance,  in the form of the potential (\ref{CPOT}). 

So far we have displayed 
the 
global $O(d,d;\mathbb{R})$ symmetry 
of the two-derivative reduced effective theory. 
The 
arguments in sec.~2 show, however, that this continuous symmetry 
is preserved by arbitrary $\alpha'$ corrections. How do we exhibit 
this symmetry to higher orders in $\alpha'$?
To first order in $\alpha'$ a natural possibility  
 is suggested by the results of Bergshoeff and de Roo~\cite{Bergshoeff:1989de}, as recently used in double field theory~\cite{Bedoya:2014pma}. 
They noted   
that the ${\cal O}(\alpha')$ Riemann-squared corrections 
can be introduced by treating the torsionful 
spin connections  
 \be\label{torspin}
  \hat\omega_{\hat\mu\, \hat a\hat b}^{(-)}(\hat e,\hat b) \ \equiv \ \hat\omega_{\hat\mu\,\hat a
  \hat b}(\hat e)-\tfrac{1}{2} \,\hat H_{\hat\mu \hat\nu\hat\rho} \, \hat e_{\hat a}{}^{\hat \nu} 
  \hat e_{\hat b}{}^{\hat \rho}\;, 
 \ee
on the same footing as the $SO(32)$ or $E_8\times E_8$ gauge fields.   
These spin connections transform 
like the gauge fields under supersymmetry
and enter the action in the 
same way:  with Chern-Simons-type modifications of the three-form field strength 
and a Yang-Mills term that for the Lorentz connection encodes a Riemann-squared term.  
The coefficient of the Lorentz Chern-Simons term, however,  
is opposite to that for the Yang-Mills Chern-Simons term.
Thus, we may simply include the ${\cal O}(\alpha')$ corrections by 
formally extending the gauge group to include the Lorentz group, 
with $\kappa$ chosen to be the negative of the Cartan-Killing metric    
 \be
   \hat\kappa_{\,\hat a\hat b,\hat c\hat d}\ = \ - \,\tfrac{1}{2}\,\alpha'\,
   \hat\eta^{}_{\,\hat a[\hat c}\,  
   \hat\eta_{\,\hat d]\hat b}\;.
 \ee 
Now our formulas of sec.~5 apply also for this case. There is,
however, an important 
subtlety:   
the definition (\ref{torspin}) means we cannot treat  
$\hat\omega^{(-)}$
as an independent gauge field.  In particular we cannot assign to it independent 
$O(d,d)$ transformations, as would follow by taking $\hat \omega^{(-)}$ to be 
part of an extended generalized metric or constrained ${\cal C}$ field. 
Rather, its $O(d,d)$ transformations to lowest order 
are fixed by those of $\hat e$ and $\hat b$, and it needs to be verified that 
these transformations are compatible. While this is very likely the case, given 
the checks performed in \cite{Bergshoeff:1995cg,Bedoya:2014pma}, 
it would be  
desirable to have a formalism in which this is manifest. 
This may also shed a new light on the double field theory  
formulations including higher derivative ${\cal O}(\alpha')$ corrections. 

\section*{Acknowledgments} 
We thank Oscar Bedoya, Eric Bergshoeff, Diego Marques, Henning Samtleben and Misha Vasiliev 
for useful discussions. The work of O.H. is supported by a DFG Heisenberg fellowship. 
The work of A.S. was
supported in part by the 
DAE project 12-R\&D-HRI-5.02-0303 and J. C. Bose fellowship of 
the Department of Science and Technology, India.
The work of B.Z. and O.H. is supported by the 
U.S. Department of Energy (DoE) under the cooperative 
research agreement DE-FG02-05ER41360.

\end{document}